\documentclass[%
 reprint,
superscriptaddress,
 amsmath,amssymb,
 aps,
 longbibliography,
]{revtex4-2}

\usepackage{graphicx}
\usepackage{dcolumn}

\usepackage{float}
\makeatletter
\let\newfloat\newfloat@ltx
\makeatother

\usepackage{algpseudocode}
\usepackage{algorithm}
\usepackage{bm}
\usepackage{color}

\DeclareMathOperator*{\argmin}{arg\,min}

\begin{document}

\preprint{APS/123-QED}

\title{Effective Dynamics of Generative Adversarial Networks}

\author{Steven Durr}
 \affiliation{Department of Physics and Astronomy, University of California Los Angeles, Los Angeles, CA 90095, USA}

\author{Youssef Mroueh}
\affiliation{IBM T. J. Watson Research Center, Yorktown Heights, NY 10598}%

\author{Yuhai Tu}%
\affiliation{IBM T. J. Watson Research Center, Yorktown Heights, NY 10598}%

\author{Shenshen Wang}
\email{shenshen@physics.ucla.edu}
 \affiliation{Department of Physics and Astronomy, University of California Los Angeles, Los Angeles, CA 90095, USA}

\date{\today}

\begin{abstract}

Generative adversarial networks (GANs) are a class of machine-learning models that use adversarial training to generate new samples with the same (potentially very complex) statistics as the training samples. 
One major form of training failure, known as mode collapse, involves the generator failing to reproduce the full diversity of modes in the target probability distribution. 
Here, we present an effective model of GAN training, which captures the learning dynamics by replacing the generator neural network with a collection of particles in the output space; particles are coupled by a universal kernel valid for certain wide neural networks and high-dimensional inputs.
The generality of our simplified model allows us to study the conditions under which mode collapse occurs. 
Indeed, experiments which vary the effective kernel of the generator reveal a mode collapse transition, the shape of which can be related to the type of discriminator through the frequency principle.
Further, we find that gradient regularizers of intermediate strengths can optimally yield convergence through critical damping of the generator dynamics. Our effective GAN model thus provides an interpretable physical framework for understanding and improving adversarial training.


\end{abstract}

\maketitle


\section{\label{sec:level1}Introduction}
In the past decade, deep generative models have proven to be an impressive tool for sampling from complex distributions. In particular, generative adversarial networks (GANs) have been used to produce realistic data, and represent a powerful framework for training generative models \cite{gan_review, style_based_gan, scene_generation, face_generation}. Consequently, understanding and improving the training of GANs is of considerable interest.

GANs comprise two neural networks: one called the \textit{generator}, $G_\theta$, and the other called the \textit{discriminator}, $D_\phi$ (parameterized by $\theta$ and $\phi$, respectively).
\begin{align}
    \text{Generator }& G_\theta: \mathbb{R}^n \rightarrow \mathbb{R}^d \\
    \text{Discriminator }& D_\phi: \mathbb{R}^d \rightarrow \mathbb{R}
\end{align}
The generator is a function which maps randomly selected points in the latent space to points in data-space.
The discriminator assigns scores to these simulated data-points, as well as to genuine samples from the data-set. During training, the discriminator's goal is to distinguish real data from simulated data (through high and low scores, respectively), while the generator's goal is to increase the score assigned to its outputs by the discriminator \cite{gan_paper, deep_learning_book, wgan, mmd_gan}.

Although GANs are both powerful and popular, they are notoriously hard to train.
The adversarial nature of the dynamics distinguishes a GAN's objective, $\mathcal{L}(\phi, \theta)$, from a standard loss function -- one that is bounded from below and which the training algorithm seeks to minimize. Rather than living at the minimum, the ideal parameter settings here are at the saddle points of the loss landscape \cite{gan_paper}:
\begin{equation}
    \theta^* = \argmin_\theta \max_\phi \mathcal{L}(\phi, \theta).
\end{equation}
Convergence to such an equilibrium is difficult to attain, as it requires a careful balancing of the two competing networks during training.

One important form of non-convergence commonly encountered during GAN training is known as \textit{mode collapse} \cite{veegan, mode_reg}. Mode collapse occurs when samples from the generator fail to capture the full diversity of modes present in the data-set. Instead, the generator's output ``collapses" as it only produces samples from relatively few of the available modes in the data distribution.

When mode collapse occurs, during training the generator will focus its distribution on a small subset of the overall data-distribution. Eventually, the discriminator learns to identify the concentrated output of the generator, at which point the generator will switch from its current specialization to another \cite{veegan, mode_reg}. The generator's output switching from mode to mode, rather than converging on the distribution as a whole, is a key symptom of mode collapse.

Many practically useful training techniques for avoiding mode collapse have been proposed, often involving modified objective functions and novel regularizers \cite{veegan, wgan, gan_paper, gan_review}.
Here, rather than constructing empirical methods for reducing mode collapse, 
we seek to understand this phenomenon from the perspective of dynamical systems, determine the physical meaning of competing factors, and derive principles to guide the training of GANs.

The dynamics of learning in neural networks have been studied in weight space \cite{feng2020neural,feng2021phases}. Here, we map GANs to an effective model in which the output of the generator network is replaced by $N$ particles in $\mathbb{R}^d$. The learning dynamics in GANs can then be studied in the output space by following the motion of the $N$ ``output" particles, which descend the loss landscape set by both the discriminator's score function and the collective state of $N$  particles. We additionally incorporate a static neural tangent kernel (NTK) -- a feature of realistic GANs using an infinite-width generator. Within our effective model, the NTK induces a dependence of the velocity of any particle on the discriminator gradient at the location of all particles.
As a result of the sampling procedure of generators and the form of common NTKs, 
we show the presence of universality within a restricted set of neural network architectures; 
many different types of infinite-width generator neural networks may lead to the same particle dynamics.

We argue that this effective model provides a simplified and interpretable framework in which to understand mode collapse.
In particular, applying this model to a low-dimensional target distribution, we show a transition from convergence to mode collapse as a function of the NTK and the relative training time. We provide a physical interpretation which explains this transition in terms of learning characteristics of the discriminator.

Finally, we use this model to study GAN regularization --  modification of the training objective in order to promote convergence. We find that when a gradient regularizer \cite{youssef_mmd_paper} is introduced, it results in a reduction of mode collapse in our model GAN.
Additionally, by sweeping over regularization strengths, we are able to observe under-regularized, over-regularized, and critically-regularized regimes. 
These regimes can be understood by analogy to the physics of a damped oscillator and its under, over, and critically damped cases.
The regularizer, which incentivizes a `smoother' generator, here plays the role of a damping term. 

\begin{figure}[H]
    \centering
    \includegraphics[width=.45\textwidth]{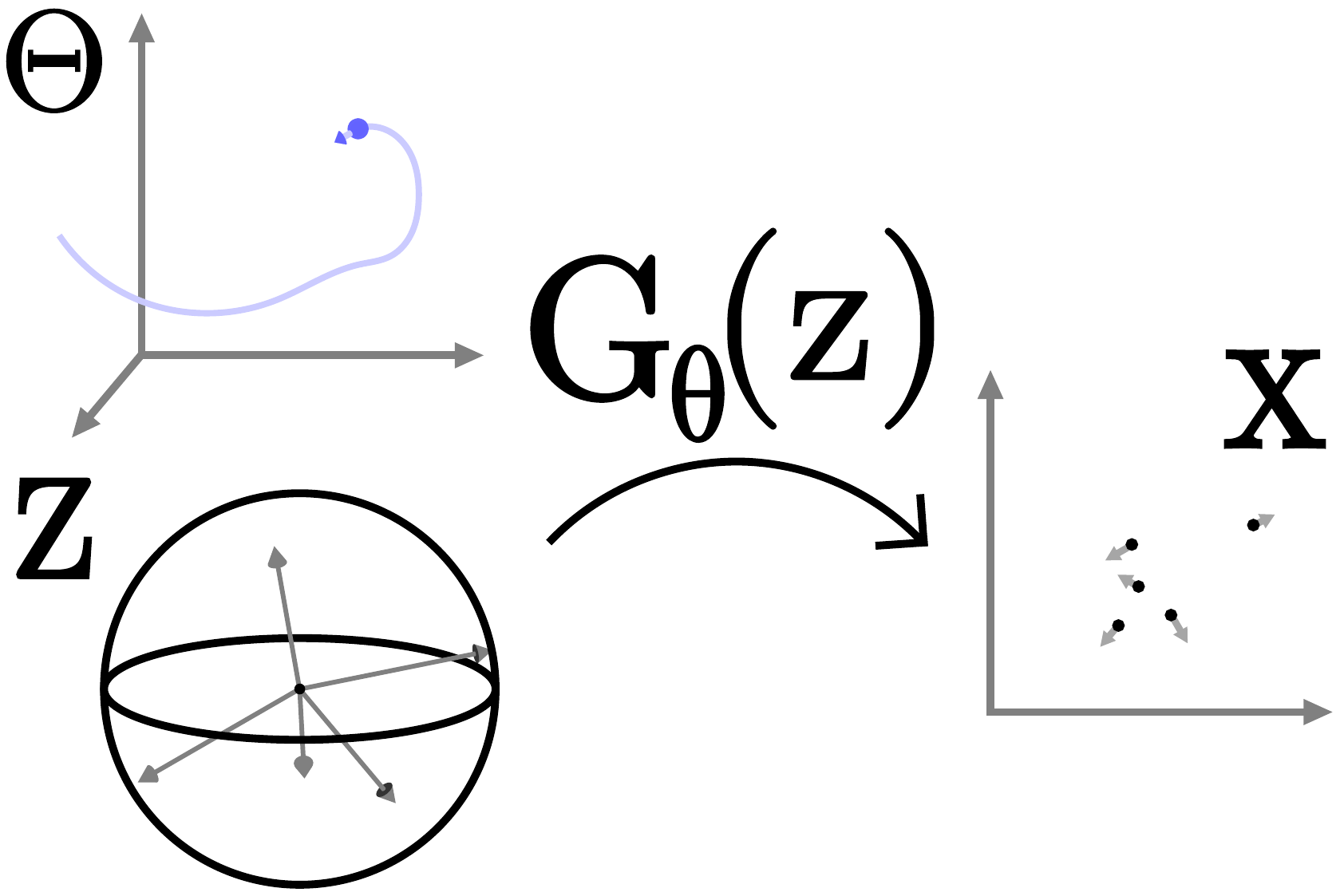}
    \caption{
    \textbf{Mapping to an effective GAN model.}
    An illustration of how the input vectors $z$ in the seed space $\mathbf{Z}$ (sampled from a high-dimensional sphere) map to particles in the data space $\mathbf{X}$. 
    During GAN training, generator parameters, $\theta_t$, evolve over time in the $\mathbf{\Theta}$ space (upper left, blue trajectory). As a result, given fixed inputs $\{ z \}$ (lower left), the set of data-space outputs, $\{X_t\}$, also evolves in time (lower right). It is the dynamics of these points in data space that our effective model directly describes.}
    \label{fig:dynamics}
\end{figure}

\section{Training and Failure}

In GANs, the generator is a neural network, $G_{\theta}$, which is fed random inputs, $z \in \mathbb{R}^n$, selected from some noise distribution $q(z)$. 
 The generator outputs, $G_{\theta}(z) \in \mathbb{R}^d$, therefore represent samples from its implicit probability distribution in data-space, $p_{\theta}(X)$.
 Conceptually, the generator and discriminator seek to minimize and maximize an objective expressing the expected difference between the data-set and the generator's outputs: 
\begin{equation}
    \mathcal{L}(\phi, \theta) = \langle D_\phi(x) \rangle_{x\sim p(x)} - \langle D_\phi(G_\theta(z)) \rangle_{z \sim q(z)}.
\end{equation}
The function $p(x)$ is the probability distribution of samples in the data-set, while $q(z)$ is the distribution from which seeds in the latent space are sampled.

In practice, however, the discriminator's objective is often modified to include regularization, restricting the magnitude of the discriminator network and promoting stability \cite{wgan, general_gan}.
Different GAN implementations exist, many with distinct objectives \cite{general_gan}. 
Here we consider objective functions of the following form \cite{orig_mmd_gan, mmd_gan, wgan, general_gan} that characterize the discriminatory power under constraints:
\begin{align}\label{eq:loss_full}
    \mathcal{L}_D &\equiv \langle D_\phi(G_\theta(z)) \rangle_{z \sim q(z)} - \langle D_\phi(x) \rangle_{x\sim p(x)} + \lambda R(D_\phi, G_\theta),\\
    \mathcal{L}_G &\equiv \langle D_\phi(x) \rangle_{x\sim p(x)} - \langle D_\phi(G_\theta(z)) \rangle_{z \sim q(z)}.
\end{align}
$\mathcal{L_D}$ and $\mathcal{L_G}$ define the objectives for the discriminator and generator, respectively, where $R(D_\phi, G_\theta)$ represents a regularizer on the discriminator, limiting its magnitude under a norm of interest (here, we use an $L_2$-norm on the discriminator weights, $\phi$) with $\lambda \geq 0$ denoting the strength of the regularizer. 

The discriminator parameters, $\phi$, evolve to maximize the expected difference between the discriminator's value on the real data and the generated data (Eq.~\ref{eq:loss_full}), while the generator parameters, $\theta$, evolve to minimize this difference.
\begin{equation}
    \dot{\phi} = -\alpha_D \frac{d \mathcal{L}_D}{d \phi}, \quad
    \dot{\theta} = -\alpha_G \frac{d \mathcal{L}_G}{d \theta}.
\end{equation}
The discriminator and generator evolution occurs at individual learning rates $\alpha_D$ and $\alpha_G$.

Practically, in neural networks, the loss function is defined using mini-batches of $N$ samples of real data and generated data, both of which are re-sampled at each training step:
\begin{align}
    \mathcal{L}_D^{(N)} &\equiv \frac{1}{N} \sum_{i=1}^N D_\phi(G_\theta(z_i)) - \frac{1}{N} \sum_{i=1}^N D_\phi(x_i) + \lambda R(D_\phi, G_{\theta}), \\
    \mathcal{L}_G^{(N)} &\equiv \frac{1}{N} \sum_{i=1}^N D_\phi(x_i) - \frac{1}{N} \sum_{i=1}^N D_\phi(G_\theta(z_i))
\end{align}
Training is performed in iterations. First, for $n_{disc.}$ steps, the discriminator is updated according to its stochastic gradient:
\begin{equation}
    \phi \xleftarrow[]{} \phi - \alpha_D \nabla_\phi \mathcal{L}_D^{(N)}.
\end{equation}
Then, for a single step, the generator is updated analogously with stochastic gradient descent:
\begin{equation}
    \theta \xleftarrow[]{} \theta - \alpha_G \nabla_\theta \mathcal{L}_G^{(N)}.
    \label{eq:generator_update}
\end{equation}
Alternating updates are repeated until convergence, or until training is halted after a large number of iterations.

Mode collapse occurs when the generator's outputs focus on a few of the available modes, rather than replicating the full data-distribution. During training, once the discriminator learns that the generator is focused at a particular mode, it assigns low scores to the data-points coming from this mode. The response of the generator is then to shift its output distribution to another mode.
Mode collapse is therefore characterized by the generator's distribution switching from mode to mode throughout training. 

\section{Generator Particles and Universality}
Rather than following the dynamics of generator parameters (Eq.~\ref{eq:generator_update}), we study instead the time evolution of generator outputs treated as particles in data-space (Fig.~\ref{fig:dynamics}), an approach applied in \cite{sobolev} and later in \cite{ntk_perspective_gans}. While each generator parameter follows its own local (stochastic) gradient, as we will show below, the dynamics of generator outputs are explicitly correlated.

With a time-dependent vector of parameters, $\theta_t$, a fixed seed $z$ maps to a point in data-space at time $t$ according to
$$
X_t = G_{\theta_t}(z).
$$
This mapping relates updates in parameter space, $d\theta_t$, to updates in data-space, $d X_t$, by
\begin{equation}\label{eq:basic_data_dynamics}
    d X_t = \frac{d G_{\theta} (z)}{d \theta}^T|_{\theta=\theta_t} \frac{d \theta_t}{d t} dt.
\end{equation}
Under gradient dynamics, the generator parameters evolve according to
\begin{equation}\label{eq:sim_gan_gen}
     \dot{\theta}_t = -\alpha_G \ \frac{d \mathcal{L}_G}{d \theta_t}=\alpha_G \ \frac{d}{d \theta_t}\langle D_\phi(G_\theta(z))\rangle_{z\sim q(z)},
\end{equation}
and so
\begin{align}
    \frac{d \theta_t}{d t} &= \alpha_G \frac{d}{d \theta} \left(\int dz' q(z')  D(G_{\theta} (z'))\right)|_{\theta=\theta_t} \\
    &= \alpha_G \int dz' q(z')  \nabla_j D(G_{\theta_t} (z')) \frac{\partial G^j_{\theta} (z')}{\partial \theta}|_{\theta=\theta_t}.
\end{align} 
To see the corresponding data-space dynamics, we plug this into Eq.~\ref{eq:basic_data_dynamics} and write
\begin{equation}
    dX_t^i = \alpha_G \ dt  \int dz' \ \Gamma^{i, j}_{\theta_t}(z, z') \nabla_j D(G_{\theta_t} (z')) q(z'), \label{eq:data_space_dynamics_1}
\end{equation}
where $i$ and $j$ index the components of the data vector $X_t$ and repeated indices are summed over. 

Moreover, we have introduced the \textit{neural tangent kernel} (NTK) \cite{jacot_ntk}, $\Gamma^{i, j}_{\theta}(z, z')$, defined by
\begin{align}
    &\Gamma_\theta: \mathbb{R}^{n} \times \mathbb{R}^{n} \rightarrow \mathbb{R}^{d \times d},\\
    &\Gamma_\theta^{i, j}(z, z') = \sum_k \frac{\partial G^i (z)}{\partial \theta_k}\frac{\partial G^j (z')}{\partial \theta_k},
\end{align}
where $n$ and $d$ are the dimensions of the inputs and the data-space, respectively, and $\theta_k$ denotes the $k$th generator parameter. 
Importantly, Eq.~\ref{eq:data_space_dynamics_1} makes clear that the NTK, $\Gamma_{\theta}$, couples the generator outputs; it specifies to what extent the dynamics of the generator particle at $X=G_\theta(z)$ is influenced by the discriminator gradients at the position $X'=G_\theta(z')$ of all other particles.

In general, NTKs evolve during training. However, for larger-width networks, the weights, $\theta_t$, will asymptotically remain in the vicinity of their initial values, $\theta_0$. The network's NTK, which involves a sum over the network's weights, changes even less -- in the infinite-width limit becoming fixed at initialization \cite{jacot_ntk, deep_learning_theory_book} (see Appendix~\ref{sec:ntk_evolution} for an example of such large-width training dynamics). In this work, we will assume that the generator is in this infinite-width regime, and enforce that the generator NTK remains fixed during training: $\Gamma_{\theta_t} = \Gamma_{\theta_0}$.

The infinite-width regime is of particular interest, as the performance of neural networks has been observed to improve as their width is increased. Additionally, in this limit, it becomes possible to derive analytical results, as certain theoretical aspects of neural networks simplify \cite{ deep_learning_theory_book, Hanin_Finite, Halverson_2021, jacot_ntk, wide_networks_are_linear}. The exact form of the infinite-width NTK can be found for particular network architectures, such as those with a ReLU or Erf activation \cite{wide_networks_are_linear}.


\subsection{Mapping to model GANs}

In generative adversarial networks, random seeds are provided to the generator by sampling from a so-called noise distribution, $q(z)$, at each iteration. Usually, this is taken to be a high-dimensional Gaussian. Noting that points from $\mathcal{N}^n(0, 1)$ are approximately on a sphere of radius $\sqrt{n}$ in $n$ dimensions \footnote{
Note that a vector selected from $\mathcal{N}^n(0, \sigma^2)$ will have an average squared length of $n \sigma^2$, and the relative standard deviation of this estimate will drop as $\sqrt{\frac{2}{n}}$.
},
we take the noise distribution as a uniform selection from a $(n-1)$-sphere.

For certain activation (most prominently, ReLU), if input seeds have a fixed magnitude, then the infinite-width NTK will be a function only of the angle between inputs: $\Gamma(z, z') = \Gamma(\varphi_{z, z'})$ \cite{wide_networks_are_linear, kernel_methods_in_deep_learning}.
Additionally, these samples selected uniformly from a high-dimensional sphere will, with high probability, be nearly orthogonal \footnote{Two elements, $z$ and $z'$ uniformly selected from an $(n-1)$-sphere of radius $\sqrt{n}$ will have a dot product obeying
$$
 \cos(\varphi_{z, z'}) \sim \mathcal{N}(0, \sigma^2 = \frac{1}{n}).
 $$
}. Therefore, given such an NTK and high-dimensional inputs, it becomes possible to estimate the distribution of NTK values within a mini-batch.

\begin{figure}[ht]
\includegraphics[width=.5\textwidth]{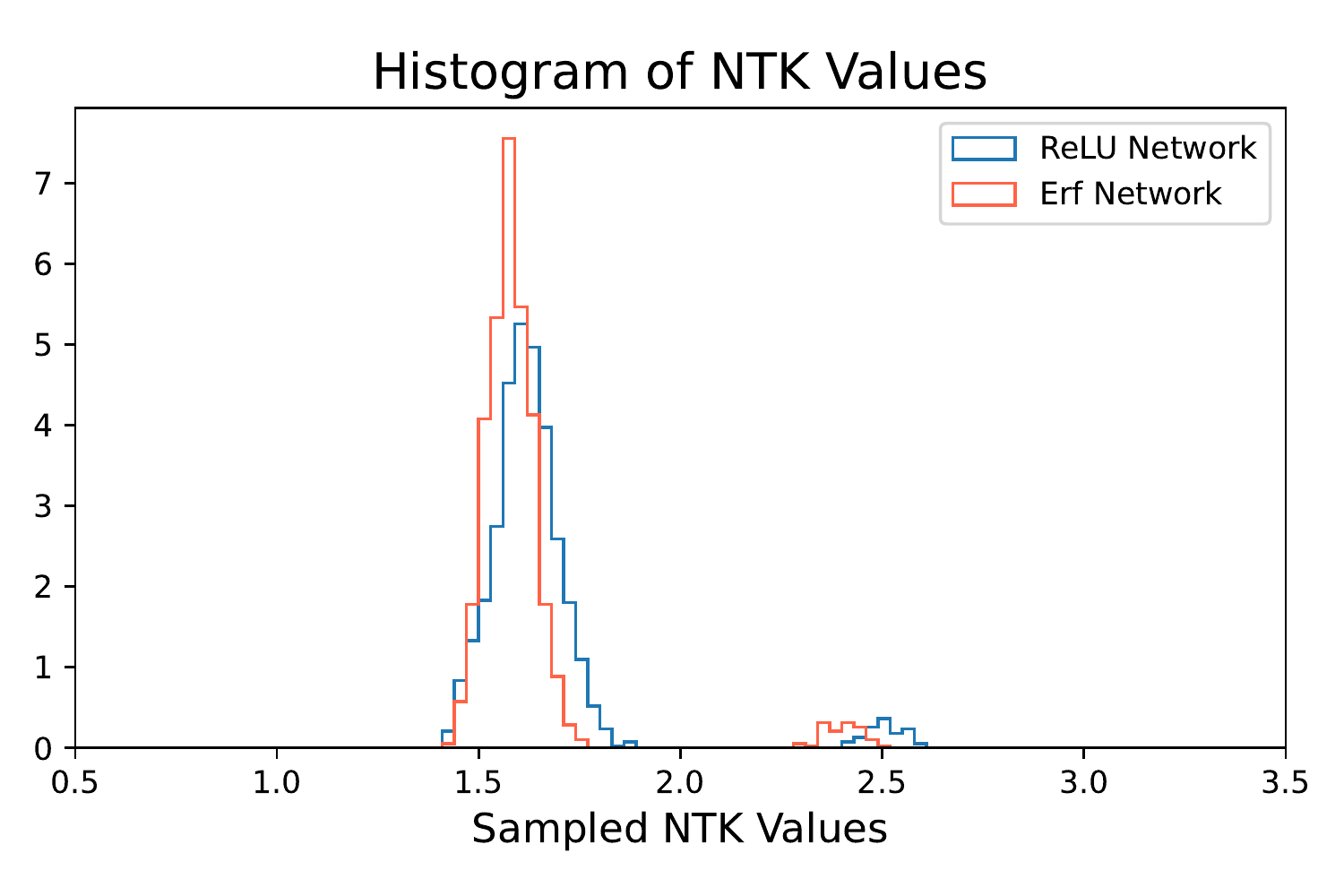}
\caption[hist_caption]{
\textbf{Universality of NTK values.}
Two distinct network architectures result in similar distributions of sampled NTK values.
50 inputs, $\{ z_i \}$, are sampled from a unit sphere in 100 dimensions. Using two untrained networks, the NTK values for all pairs of inputs are computed. 
The red histogram is obtained using a single hidden layer network (width 2048) with an Erf activation, while for the blue histogram a ReLU activation is used. Zero-mean normal distributions with distinct variances are used to initialize two networks' weights and biases.
Despite their differences, the two networks' NTK values are approximately characterized by the same two numbers: $\Gamma(\varphi_{z, z'} \approx \pi/2)$ for distinct inputs and $\Gamma(\varphi_{z, z} = 0)$ for pairs of the same input.
}
\label{fig:NTK_distribution}
\end{figure}

Using these observations, we propose a simplification of the GAN training protocol. Within our simplified model, we take the generator to be of large width with a static NTK. The noise distribution, $q(z)$, is taken to be a uniform distribution over a high-dimensional sphere. 
Finally, (as in wide ReLU networks with inputs of fixed magnitude), we take the NTK to be a function of the dot-product of inputs only. 

Our first assumption fixes the NTK at initialization \cite{wide_networks_are_linear, jacot_ntk}. The latter two concentrate the pairwise NTK values, obtained using one sample of inputs $\{z \}$, to two characteristic numbers $g_1$ and $g_2$. 
The first number, $g_1 \equiv \Gamma(\varphi_{z, z} = 0)$, corresponds to evaluations involving the same point, and the second, $g_2 \equiv \Gamma(\varphi_{z, z'} = \pi/2)$, corresponds to pairs of distinct points chosen from the high-dimensional latent space.
The values of $g_1$ and $g_2$ are determined by the architecture of the network, but can also be modified by, for instance, the use of batch-normalization \cite{freeze_chaos_ntk, batchnorm}.

The fact that an entire generator neural network, with its activation functions and individual weight and bias distributions, can be to an extent characterized by just two numbers, suggests a sort of universality within this particular set of neural network architectures. Many different generator neural network architectures may be mapped on to the same system, parameterized only by ($g_1$, $g_2$). This universality can be observed in Fig.~\ref{fig:NTK_distribution}, in which two distinct networks (one with ReLU activation, the other with Erf, both using a single hidden layer with 2048 units \footnote{
The ReLU and Erf networks have respective weights sampled from $\mathcal{N}(\mu=0, \sigma^2 \approx 0.42 \text{ and } 1.18)$, and respective biases sampled from $\mathcal{N}(\mu=0, \sigma^2 \approx 1.17 \text{ and } 11.67)$.}) are observed to have very similar pairwise NTK values across a sample from a unit sphere in 100 dimensions. 

Additionally, we consider a restricted version of GAN training. 
Rather than re-sampling from the noise distribution (i.e. taking a new mini-batch from $q(z)$) at each training iteration, we instead train by effectively using one fixed set of $N$ generator inputs, $\{z\}$. 

Based on these simplifications, we propose a coarse-grained NTK of the form
\begin{equation}\label{eq:general_ntk}
    \Gamma^{i, j}(z, z') = \delta_{i, j}\left(g_1 \delta_{z, z'} + g_2 (1-\delta_{z, z'})\right).
\end{equation}
This NTK is static throughout training, 
and its two constant values, for diagonal ($g_1$) and off-diagonal ($g_2$) entries, characterize the NTK values for pairs of identical and distinct points, respectively.

This NTK allows us to further simplify the effective model -- ignoring the latent space entirely, and instead explicitly correlating particles in data-space:
\begin{equation}\label{eq:general_ntk_2}
    \Gamma^{i, j}_{a, b} = \delta_{i, j}\left(g_1 \delta_{a, b} + g_2 (1-\delta_{a, b})\right).
\end{equation}
Here, $a$ and $b$ index particles, while $i$ and $j$ index components in data-space. Out front, $\delta_{i, j}$ can be understood as implying a lack of correlation between the gradients of output degrees of freedom of an infinite-width neural network. 
$g_1$ and $g_2$ set the degree to which the discriminator gradient at local and distinct points, respectively, contribute to a generator particle's velocity. 

Using this effective NTK, we can model the dynamics of data points in output-space by dynamics of coupled generator particles
\begin{equation}\label{eq:data_space_dynamics_2}
    \frac{d X^i_a}{dt} = \frac{\alpha_G}{N} \sum^{N, d}_{b, j} \Gamma^{i, j}_{a, b} \nabla_j D(X_b).
\end{equation}
These dynamics are reminiscent of flocking behavior, in which local velocities are found through a spatial average \cite{flock}. Here, however, the average is not over velocities, but over discriminator gradients. Additionally, the average is taken over all particles, rather than over a local region.


\subsection{Multi-Modal Target}
We now proceed with our simplified GAN training protocol, replacing the generator network with a collection of $N$ particles in data-space, and using the generator update rule of Eq.~(\ref{eq:data_space_dynamics_2}) rather than that of Eq.~(\ref{eq:data_space_dynamics_1}). 

As a case study, we consider a canonical two-dimensional problem of training a GAN on a distribution of 8 Gaussians arranged in a circle of radius 2, each having a standard deviation 0.02. Since each Gaussian can naturally represent a distinct mode, this data distribution is used throughout GAN literature as a toy data-set for observing mode collapse \cite{veegan, mode_reg, youssef_mmd_paper}. Mode collapse in this context would correspond to a generator whose outputs are focused on one, or a subset, of the eight Gaussians. During training, mode collapse would cause the outputs to oscillate between distinct modes, without splitting to cover all eight.

The generator particles are taken to be 2000 parameterized points in the plane, initialized as a Gaussian distribution with $\sigma=0.5$, while the discriminator is a ReLU network with 4 hidden layers of width 512 \footnote{weights are initialized using a Glorot uniform distribution \cite{glorot}, and biases are initialized at zero}.
The discriminator parameters, $\phi$, and the generator points, $X_a$, are both updated during training according to their objective functions, following training routine described in Algorithm \ref{alg:gan_training_2}.

\begin{figure}[ht]
    \centering
    \includegraphics[width=.5\textwidth]{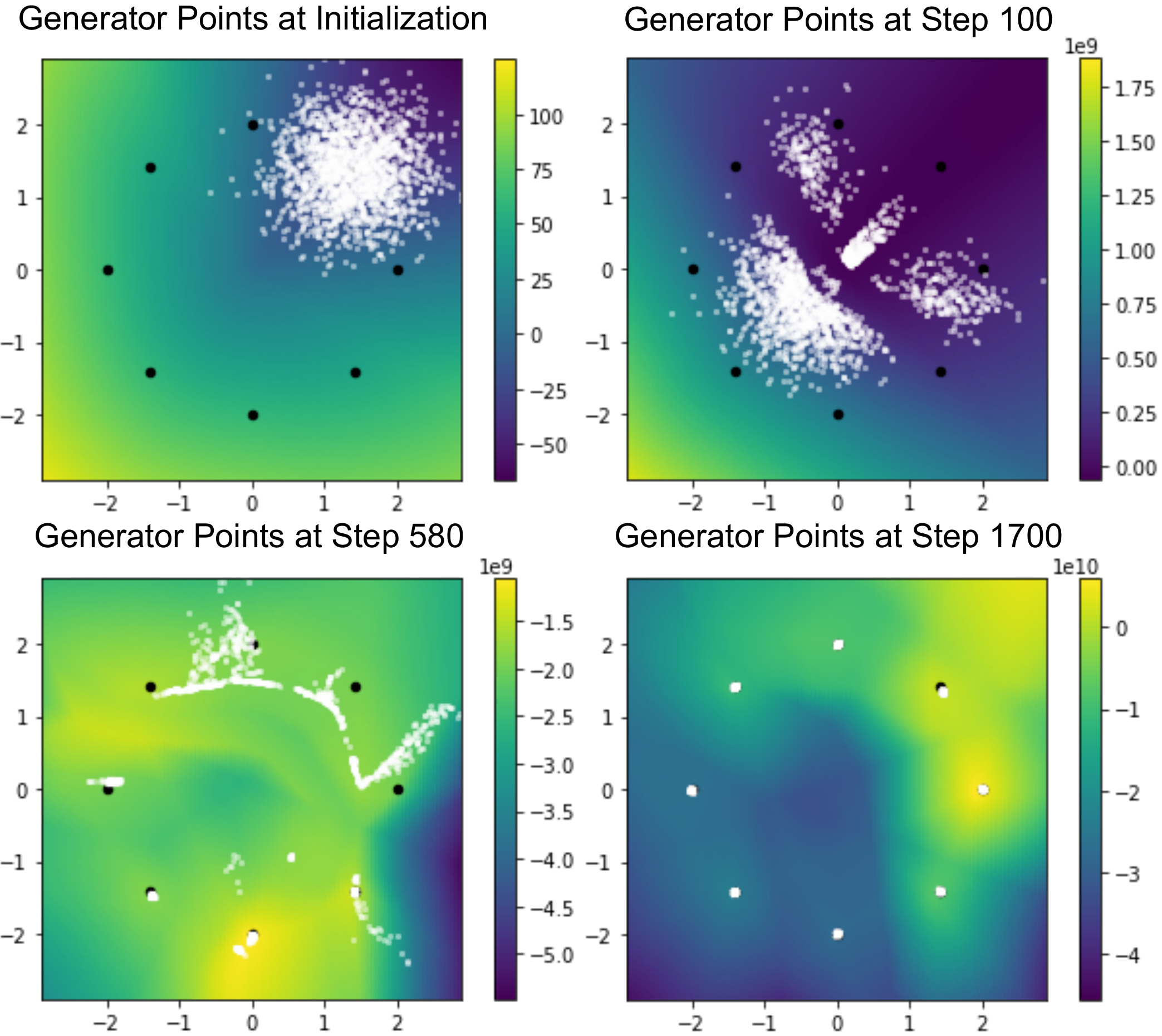}
    \caption
    {\textbf{Convergence of the model GAN dynamics under a diagonal NTK ($g_2 = 0$)}. The generator particles are shown in white, while the discriminator values are color-coded, with higher values shown in brighter colors. Because the NTK is diagonal, the particles' velocities are not explicitly correlated, and they ascend their local gradients (from darker to brighter colors), $\dot{X} \propto \nabla D(X)$. Meanwhile, the discriminator attempts to maximize the difference between its expectation on the data (black points) and generator particles (in white). Initially, the discriminator assigns a low value to the cluster's location. Consequently, the cluster rapidly splits apart (step 100), with each point following the local discriminator gradient.
     The combined adversarial dynamics are seen to result in convergence to the eight modes.}
    \label{fig:no_ntk_2d}
\end{figure}

\begin{figure}[ht]
    \centering
    \includegraphics[width=.5\textwidth]{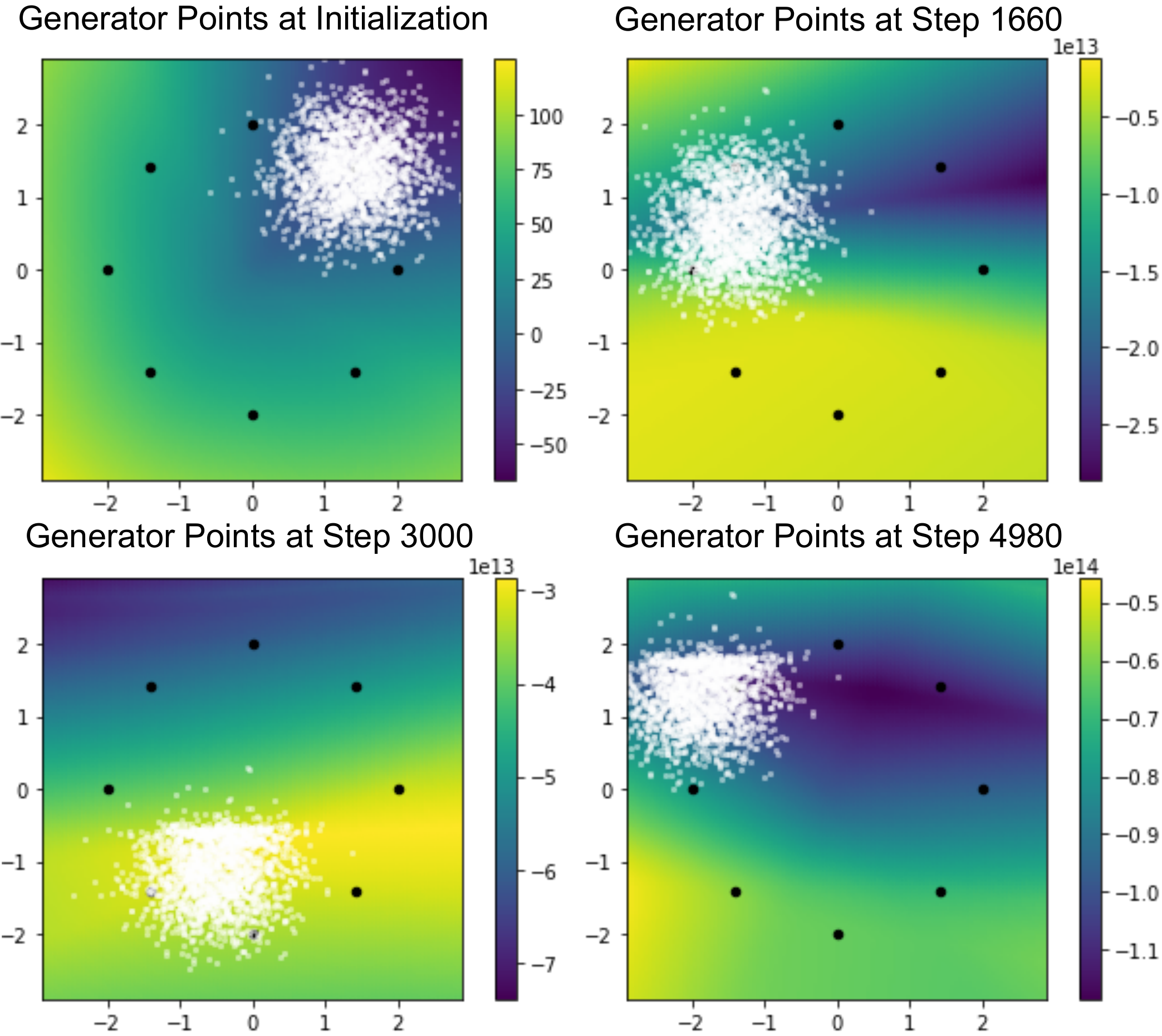}
    \caption{
    \textbf{Model GAN dynamics exhibit mode collapse under all-to-all coupling by a non-diagonal NTK ($g_2 / g_1 = 1/5$).} 
    As in Fig.~\ref{fig:no_ntk_2d}, dynamics are depicted over time.
    During training, the discriminator seeks to assign higher values (brighter colors) to real data-points (the eight points in black), and lower values (darker colors) to generated points (white dots).
    As had occurred in Fig.~\ref{fig:no_ntk_2d}, the discriminator initially places its minimum at the position of the generator particle cluster.
    Now, however, due to correlations in particle velocity, the cluster no longer splits apart. Instead, it shifts away from discriminator minima before splitting can happen.
    As a result, the cluster of generator particles switches from mode to mode, as the discriminator attempts to `catch up' -- a behavior indicative of mode collapse. }
    \label{fig:with_ntk_2d}
\end{figure}

In Figures \ref{fig:no_ntk_2d} and \ref{fig:with_ntk_2d},
 we show time slices of the training progress. The generator particles are shown in white on a heat-map of the discriminator values.
We begin by running an experiment using a diagonal NTK ($g_2 / g_1 = 0$ \footnote{
Noting the dynamics described in Eq.~(\ref{eq:data_space_dynamics_1}), we normalize the generator's dynamics by the particle number, setting $g_1=2000$, $g_2=0$ so that $\frac{1}{N}\sum^N_{a, b}\Gamma(X_a, X_b) \nabla D(X_a) = \nabla D(X_b)$.
}). 
In this case, the generator particles independently ascend the local gradient of the discriminator: $\dot{X}^i_a \propto \nabla_i D(X_a).$ Visually (Fig.~\ref{fig:no_ntk_2d}), this corresponds to each particle (in white) drifting up the color gradient (taking steps towards lighter regions). Meanwhile, the discriminator modifies its parameters to increase the difference between the expectation on the real data and the generator particles -- assigning higher values (brighter colors) to the eight data points in black, and lower values (darker colors) to the particles in white. 

We observe the result of this dynamic in Fig.~\ref{fig:no_ntk_2d}. Initially, the discriminator assigns low values to the cluster of particles. However, the initial cloud of particles rapidly splits apart, and the adversarial dynamic results in informative gradients being passed to the generator particles, which quickly converge to the full multi-modal distribution.

In a second experiment, we begin with the same initialization, but instead use an NTK satisfying $g_2 / g_1 = 1/5$. Due to the non-trivial off-diagonal terms of the NTK, the velocities of the particles are correlated. As is shown in Fig.~\ref{fig:with_ntk_2d}, generator particles (in white) no longer split apart. Instead, they stay together as the entire cluster shifts from mode to mode indefinitely. 
The discriminator repeatedly attempts to assign low values (darker colors) to the generator particle cluster's spatial region. 

This behavior is a key signature of mode collapse, and suggests an understanding of this phenomenon through the lens of our model. 
For the remainder of this work, we will identify the observed failure of convergence and switching between modes with mode collapse. 
By varying $g_2 / g_1$, we will probe the onset of this failure mode, and investigate what training algorithms and discriminator characteristics would lead to improved performance.

\begin{algorithm}
\begin{algorithmic}
\For{iteration number}
    \For{$n_{disc.}$}
        \State $\bullet$ Sample $N$ data-points, $\{ x_i \}$, from the 8-Gaussian distribution.

        \State$\bullet$  Compute
        $$
        \mathcal{L}_D^{(N)} =  \frac{1}{N} \sum_{a=1}^N D_\phi(X_a) - \frac{1}{N} \sum_{l=1}^N D_\phi(x_l) + \frac{\lambda}{2}\sum_k \phi_k^2
        $$
        \State and update discriminator parameters by descending its stochastic gradient
        $$
        \phi \gets \phi - \alpha_D \nabla_\phi \mathcal{L}_D^{(N)}
        $$
    \EndFor
    \State $\bullet$ update $X_a$ according to Eq.~(\ref{eq:data_space_dynamics_2})
    $$
    X_a^i \gets X_a^i + \alpha_G \ \frac{1}{N} \sum^{N, d}_{b, j} \Gamma_{a, b}^{i, j} \nabla_j D(X_b)
    $$
\EndFor
\end{algorithmic}
\caption{The coarse-grained, $(g_1, g_2)$ GAN training algorithm.}
\label{alg:gan_training_2}
\end{algorithm}

\section{Model GAN Experiments | \newline The Mode Collapse Transition}
\label{sec:mode_experiments}

We have observed that the ratio $g_2/g_1$ may be increased to induce mode collapse. Apart from the architecture of the discriminator network, the remaining adjustable parameters in Algorithm \ref{alg:gan_training_2} concern the relative training dynamics of the discriminator and generator. The parameters $\alpha_D$ and $\alpha_G$ control the step-size of the discriminator and generator, respectively, while $n_{disc.}$ tunes the number of discriminator steps taken for each generator step.

We will therefore vary these parameters to examine the relationship between $g_2 / g_1$ and the discriminator's dynamics.
The latter can be varied in two ways: by modifying the learning rate $\alpha_D$, or by modifying the value of $n_{disc.}$ used in the algorithm. Here, we show the result of modifying $n_{disc.}$, leaving $\alpha_D$ experiments (which produce similar results) to Appendix~\ref{app:exp_details}.

To characterize whether, at a given time, generator particles have converged or collapsed to a single mode, we define a metric based on the entropy of the distribution. Letting $P_i$ be the fraction of particles for which the $i^{th}$ mode is the nearest,
 we define the following:
\begin{equation}\label{eq:mode_collapse_metric}
\text{Mode Collapse Metric} = \log(8) + \sum_i P_i \log(P_i).
\end{equation}
Note that $P_i = 1/8$, $i=1,2,\cdots8$, would give a complete mode coverage with an even split, and have a value of $0$. On the other hand, $P_1 = 1$, $P_{i>1}=0$ would correspond to all generator points being nearest to a single mode, giving a mode collapse metric value of $\log 8$.

To further characterize the quality of convergence of the generator particles, we can compute the average log-likelihood, $\frac{1}{N}\sum_a \log(p(X_a))$, where $p(X)$ is the probability density of the multi-modal Gaussian distribution.  
The combination of these two metrics (mode collapse and log-likelihood) indicates whether the generator points have both avoided mode collapse and successfully converged to the modes of the distribution. 

\subsection{GAN Setup}

To maximize the interpretability of our results, we employ a simpler discriminator with a single wide hidden layer (2048 units):
\begin{equation}
    D(x) = \sqrt{\frac{2}{\text{width}}} a_i \sigma(w^j_i x^j + b_i).
\end{equation}
 Details of the initialization can be found in Appendix~\ref{app:exp_details}.
The activation function is set to ReLU, $\sigma(x) = \max(0, x)$. Experiments employing a Tanh activation were also performed and the results can be found in Appendix~\ref{app:fprince_support}.
The target data distribution is again taken to be the eight Gaussians. 
A total of 200 generator particles are initialized at $(\frac{1}{\sqrt{2}}, \frac{1}{\sqrt{2}})$ with a standard deviation of $0.1$.

The discriminator loss function is defined as
\begin{multline}
  \mathcal{L}_D = \langle D(X) \rangle_{\text{gen.}} - \langle D(x) \rangle_\text{target}\\
  + \frac{1}{\text{width}} \left(\sum_{i, j} (w_i^j)^2 + \sum_i a_i^2 + \sum_i b_i^2 \right),
\end{multline}
where $\langle D(X) \rangle_{\text{gen.}}$ is the expectation of the discriminator on the generator distribution, $\frac{1}{N}\sum_a D(X_a)$, while $ \langle D(x) \rangle_\text{target}$ is its expectation on the data distribution (the eight Gaussians). The remaining terms represent an $L_2$-regularizer on the weights, placing an overall restriction on the discriminator.

Following Eq.~\ref{eq:data_space_dynamics_2}, particle velocities are given by
\begin{align}
    \frac{d X^i_a}{dt} &= \frac{\alpha_G}{N} \sum_{b, j}^{N, d} \Gamma^{i, j}_{a, b} \nabla_j D(X_{b}) \\
    &= \alpha_G\left( \frac{g_1-g_2}{N} \nabla_i D(X_{a}) + g_2 \langle \nabla_i D(X)\rangle \right).
    \label{eq:competition_eq}
\end{align}
Here the angular bracket indicates an average over all generator particles.
Hence, each particle, at position $X_a$, experiences a competition between the mean discriminator gradient over the ensemble, $g_2 \langle \nabla D(X)\rangle$, and the contribution from its local gradient, $\left( \frac{g_1 - g_2}{N}\right) \nabla D(X_a)$.

The entire system is trained using Algorithm \ref{alg:gan_training_2}.

\begin{figure}[t]
\includegraphics[width=.5\textwidth]{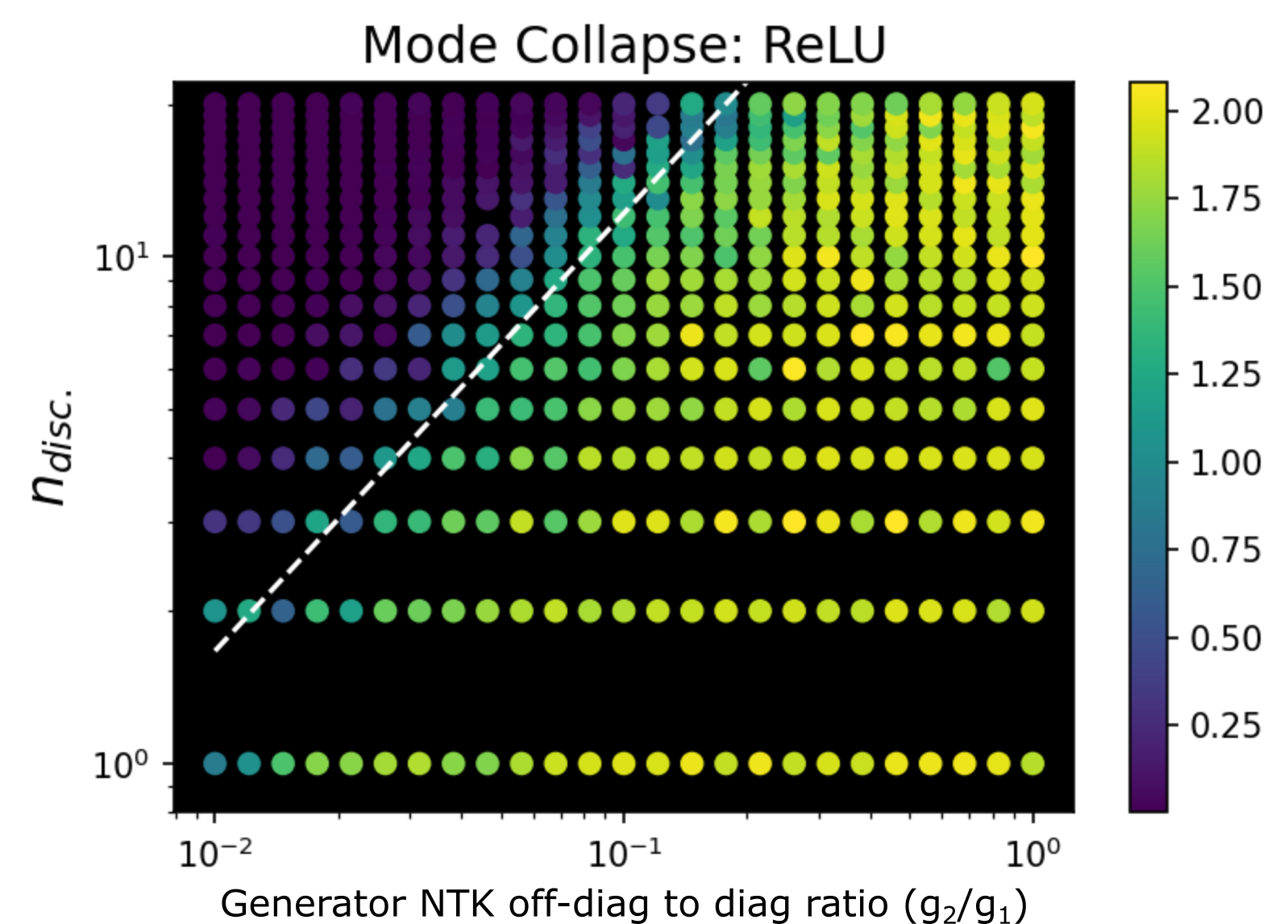}
\caption{
\textbf{Phase diagram of the transition to mode collapse using a ReLU discriminator.}
Mode collapse metric values are shown as a function of generator NTK and discriminator training rate. 
Brighter points indicate mode collapse while darker points correspond to an even distribution of generator particles over the target data.
The x-axis gives the value of $g_2 / g_1$ for a given experiment, while the y-axis indicates the number of training steps the discriminator takes at each iteration.
As $g_2 / g_1$ is increased, the discriminator requires more `time' (a greater $n_{disc.}$ value) in order to shift the training from mode collapse (bright points) to convergence (darker points).
Experimental results are taken after 5000 training iterations, and data is time-averaged, representing the mean of results taken at $5000 \pm n\times 20$, with $n = 0, 1, 2, 3$. A line is fit to $(1/2) \log 8$, indicating a power-law boundary.}
\label{fig:relu_data}
\end{figure}

\subsection{Results and Interpretation}
\label{interpretation}

We run the model GAN training algorithm for each ($g_2 / g_1$, $n_{disc.}$) pair considered.
After training for a fixed number of iterations
, and computing the mode collapse metric (Eq.~\ref{eq:mode_collapse_metric}) for all pairs, we can observe a clear transition from convergence (blue data-points) to mode collapse (yellow data-points), as shown in Fig~\ref{fig:relu_data}. In the following, we provide a heuristic argument to explain the observed characteristics of the mode-collapse transition.



During training, the discriminator seeks to maximize the difference between its expectation on the training data and on the generator distribution. 
Suppose to this end, the discriminator has formed its minimum within a region (cluster) of generator particles. 
 According to the equation of motion for the generator particles (Eq. 26), if the term involving local gradients dominates over $g_2 \langle \nabla D(X) \rangle$, then the particles in this cluster `split apart'; each particle follows its own local gradient, regardless of the location of the discriminator minimum within the region. As a result,  the generator particle cluster, which corresponds to mode collapse, can be split and the full targeted distribution can be recovered. 

\begin{figure}[ht]
\includegraphics[width=.48\textwidth]{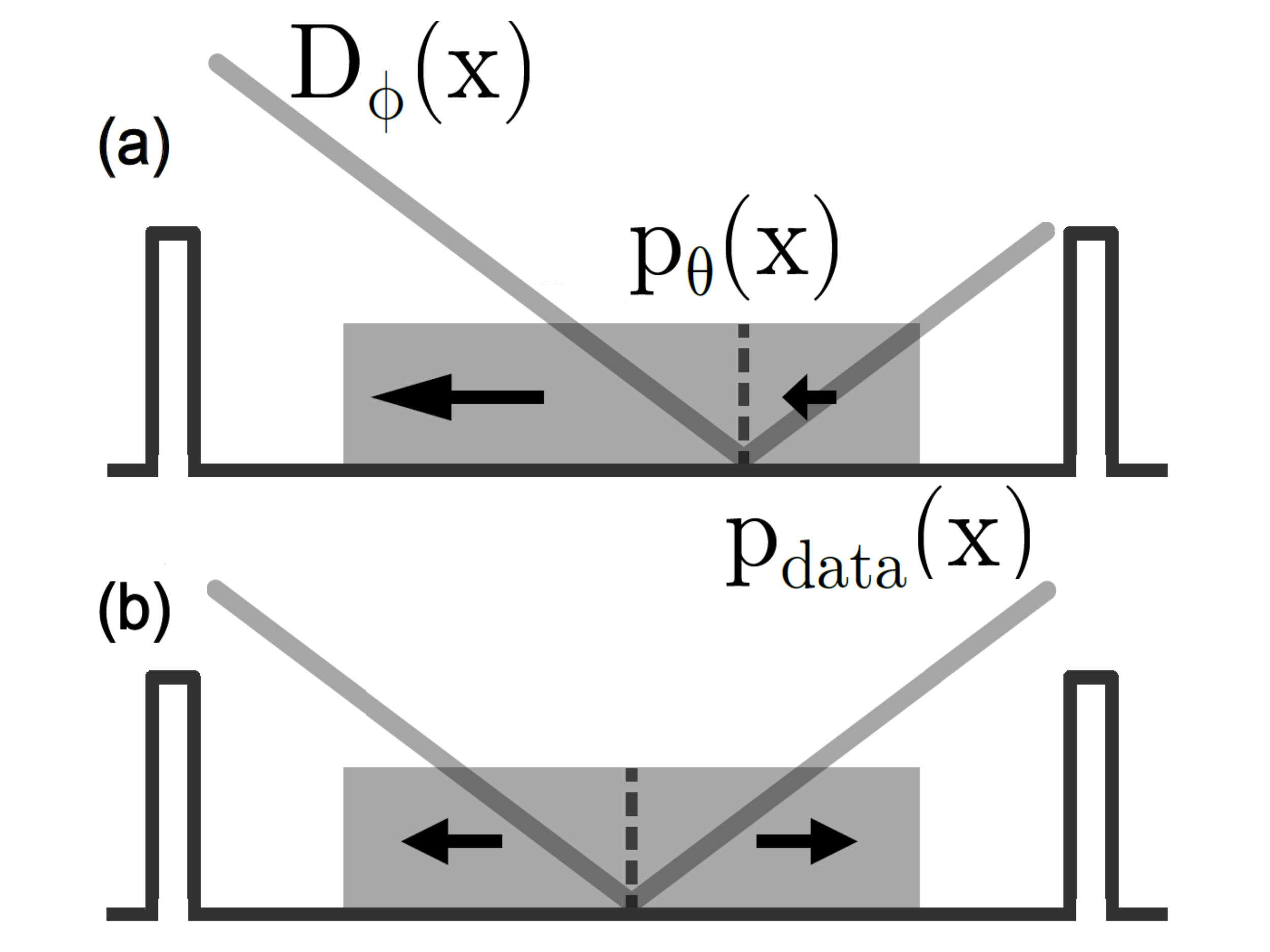}
\caption{
\textbf{Increased discriminator precision may result in splitting of a cluster of generator particles}. 
A cluster of generator particles is depicted as a uniform shaded region (with distribution $p_\theta$) between two modes of the target distribution ($p_\text{data}$). The value of the discriminator is shown as a simple function of the form $D_\phi(x) \equiv a |x-b|$. Arrows indicate the local velocities of particles on either side of the discriminator minimum. (a) Imprecision of the discriminator leads to a sub-minimal expectation, $\langle D(X) \rangle$, and a significant $\langle \nabla D(X) \rangle$ value, causing all particles to slip to the left due to all-to-all coupling through the NTK. (b) The discriminator has found the precise minimum at the center of the distribution, which kills off average gradients and allows particles to ascend their local slopes; the cluster is thus splitting apart.
}
\label{fig:slip_away}
\end{figure}

However, for sufficiently large $g_2$, the term $g_2 \langle \nabla D(X) \rangle$ may dominate. Now the location of the discriminator's minimum within the region becomes important, as it may determine both the magnitude and direction of $\langle \nabla D(X) \rangle$. As is depicted in Fig.~\ref{fig:slip_away}, if the discriminator obtains a minimum far from the center of a cluster, $|\langle \nabla D(X) \rangle|$ would become non-negligible, leading to an onset of instability, and causing the entire group of generator particles to ``slip away" -- including those on the opposing slope (panel a). In contrast, for a minimum closer to the cluster's center, the mean gradient experienced by the particle cloud becomes sufficiently small to allow the cluster to ``split apart" (panel b).

\begin{figure}[t]
\includegraphics[width=.5\textwidth]{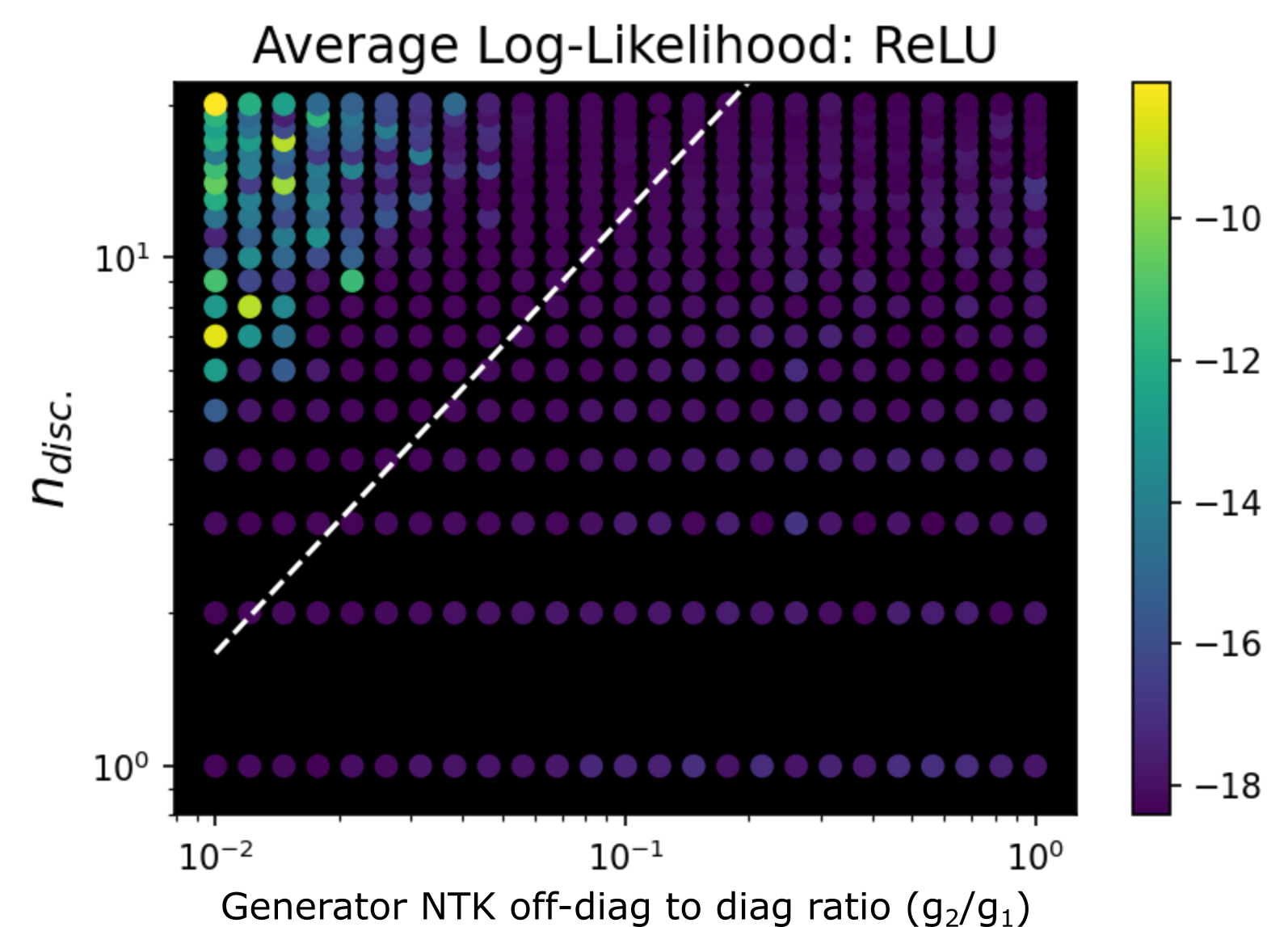}

\caption{
\textbf{Precise convergence occurs well above the mode-collapse transition.}
The average log-likelihood of generator particles is depicted following the same procedure as in Fig.~\ref{fig:relu_data}.
The power-law transition of Fig.~\ref{fig:relu_data} is shown as a reference; only sufficiently far above this boundary would particles converge precisely to the eight modes.
}
\label{fig:relu_loglikelihood_data}
\end{figure}

In this way, the discriminator's precision in minimizing its value over a cluster of generator particles influences its ability to split apart the cluster. More spatially precise discriminators may yield smaller values of $|\langle \nabla D(X) \rangle|$, allowing local gradients to dominate particle dynamics.

The mode-collapse data resulting from using a ReLU discriminator is shown in Fig.~\ref{fig:relu_data} on a log-log scale. A dashed white line emphasizes a visible power-law boundary separating mode collapse from convergence. Examples of generator particle distributions sampled across this transition can be found in Appendix~\ref{app:sample_dists}.
Fig.~\ref{fig:relu_loglikelihood_data} plots the log-likelihood data and shows that only sufficiently far above the transition boundary would particles converge precisely to the target modes.

This power-law behavior matches another feature of wide ReLU networks, referred to as the frequency principle \cite{frequency_vs_training, convergence_vs_freq, linear_f_principle}. As networks learn, they tend to first learn lower-frequency functions, before including higher-frequency contributions. This behavior thus sets a rate $\gamma(k)$ at which a network can learn a feature of spatial frequency $k$.
For example, within wide ReLU networks, $\gamma(k)$
is expected to be power law,
whereas for wide Tanh networks, an exponential $\gamma(k)$ is predicted \cite{linear_f_principle}.

If we identify spatial features of size $1/k$ as having a dominant spatial frequency of $k$, then simple arguments suggest (Appendix~\ref{section:precision_scaling_argument}) 
that in order to split a cluster, the maximum allowable spatial imprecision falls with increasing $g_2$ as $(g_1 - g_2)/g_2$.
This indicates that to break apart such a distribution, we require the discriminator to learn a feature with spatial \textit{frequency} proportional to $g_2 / (g_1 - g_2)$; for $g_2 \ll g_1$, roughly, $k \sim g_2 / g_1$.

Assuming that the discriminator has a frequency-dependent learning rate, $\gamma(k)$, then the time required to learn such a feature scales as $T \sim 1/\gamma(k)$. The necessary discriminator steps, $n_{disc}$ (and learning rate, $\alpha_D$), to overcome mode collapse would then scale as $1/\gamma(k)$. Interestingly, experiments involving Tanh discriminators (and other types) seem to support such a conclusion (Appendix~\ref{app:fprince_support}).
As such, the frequency principle suggests a connection between the relevant time-scales and length-scales of the experiment's learning objective.

\section{Critical Regularization }
\label{sec:critreg}

Various regularization techniques have been applied to the problem of mode-collapse avoidance \cite{veegan, mode_reg, youssef_mmd_paper}. 
Here, we demonstrate that even in our simplified model GAN, the effect of regularization on reducing mode collapse can be observed. 


Following \cite{youssef_mmd_paper}, we introduce a gradient regularizer 
\begin{equation}
    \beta ||\nabla_\theta \langle D(G_\theta (z))\rangle ||^2 /2
    \label{eq:reg1}
\end{equation}
into the discriminator's loss function during training.
Since the velocities of generator parameters are driven by local gradients of the discriminator, this term is analogous to the kinetic energy of these parameters. This regularization term penalizes sharp gradients and encourages generators to take smoother paths to the target distribution. The effect of the regularizer on the GAN system can be viewed in analogy to a damping term in physics (a connection made explicit in Appendix~\ref{app:dirac_gan_reg}), with oscillations from mode to mode corresponding to an under-damped regime, slow convergence to all available modes corresponding to over-damping, and a most efficient convergence corresponding to critical-damping. Using this analogy as a conceptual starting point, we can sweep over $\beta$ to identify a regime of ``critical regularization".

Despite the fact that our model GAN setup does not have any reference to generator parameters (or a generator network), we may still incorporate such a term into training via the effective NTK:
\begin{multline*}
    ||\nabla_\theta \langle D(G_\theta (z))\rangle ||^2 = \\
    \frac{1}{N^2}\sum_{z, \ z' \sim q(z)} \nabla_i D(G_\theta (z)) \nabla_\theta G^i_\theta (z) \nabla_\theta G^j_\theta (z') \nabla_j D(G_\theta (z')) \\
    = \frac{1}{N^2}\sum_{z, \ z' \sim q(z)} \nabla_{i} D(G(z)) \Gamma_\theta^{i, j}(z, z') \nabla_{j} D(G(z')).
\end{multline*}
This effective form immediately allows us to apply the regularizer within the model GAN by including the following term in Eq.~\ref{eq:loss_full}:
\begin{equation}
    \frac{\beta}{N^2} \sum_{a, b} \nabla_{i} D(X_a) \Gamma^{i, j}_{a, b} \nabla_{j} D(X_b).
    \label{eq:reg}
\end{equation}

\begin{figure}[ht]
    \centering
    \includegraphics[width=.5\textwidth]{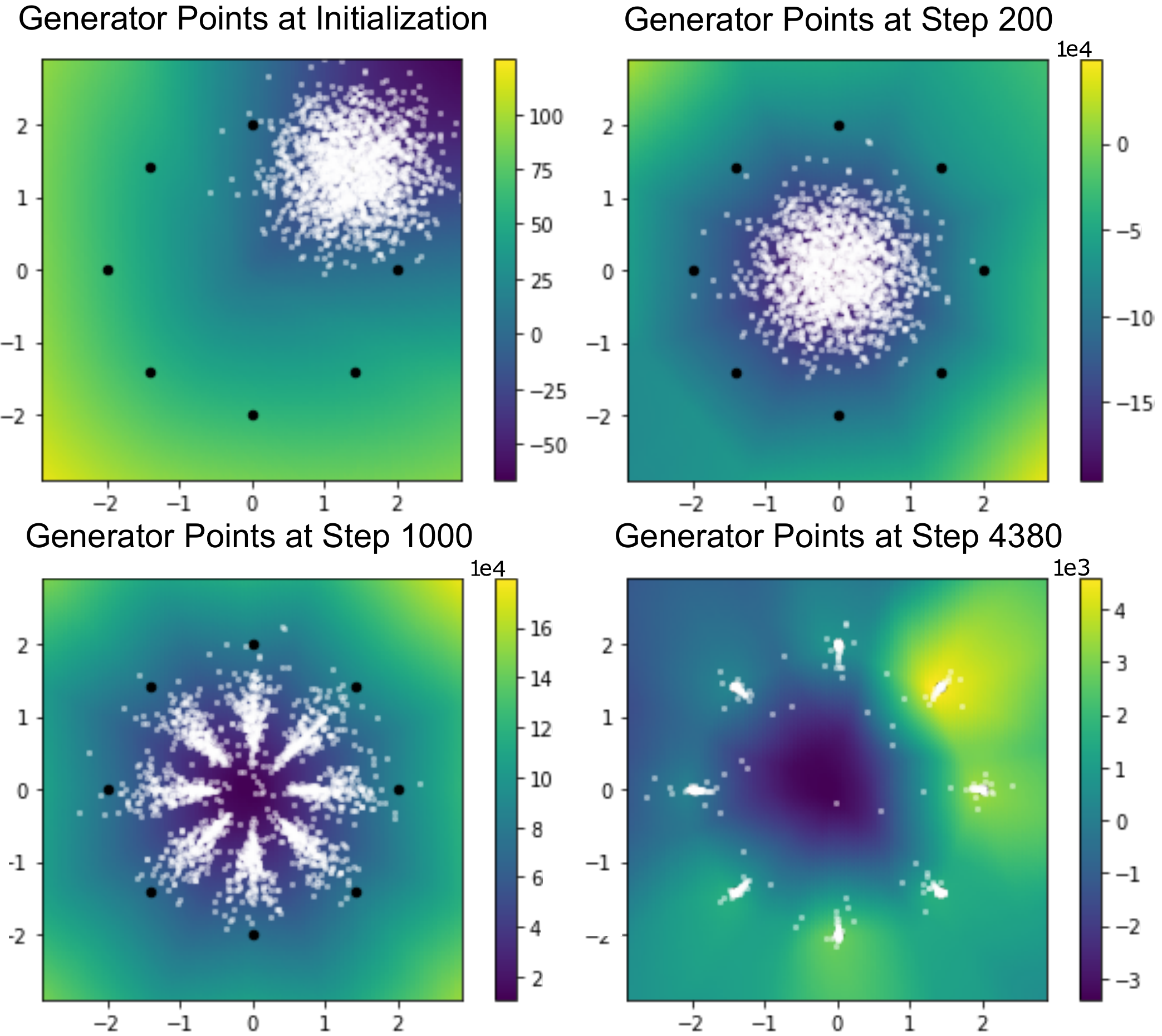}
    \caption{
    \textbf{Model GAN dynamics with significant off-diagonal NTK values ($g_2 / g_1 = 1/5$) converge under regularization.}
    Model GAN dynamics is shown over time, taking $g_2/g_1 = 1/5$ and including a regularizer (Eq.~\ref{eq:reg}, with $\beta=100$). Despite that without the regularizer the system oscillates from mode to mode (Fig.~\ref{fig:with_ntk_2d}), now particles converge evenly and steadily to the eight modes.}
    \label{fig:reg_dyn_scatters}
\end{figure}

We can understand the effect of such a regularizer by expressing (\ref{eq:reg}) in the form,
\begin{equation}
    \beta \left( g_2 |\langle \nabla D(X)\rangle|^2 + \frac{(g_1 - g_2)}{N}\langle |\nabla D(X)|^2 \rangle \right).
\end{equation}
The second term discourages sharp gradients from being provided to generator particles, leading to smoother paths to convergence. 
The first term, directly proportional to $g_2$, can be seen to discourage the presence of large mean gradients over the ensemble of generator particles. 
Incorporating this into the same setup used to produce mode collapse in Fig.~\ref{fig:with_ntk_2d} and repeating the procedure \footnote{Here, we set $\beta = 100$}, we now observe convergence instead (Fig. \ref{fig:reg_dyn_scatters}).

\begin{figure}[ht]
    \centering
    \includegraphics[width=.5\textwidth]{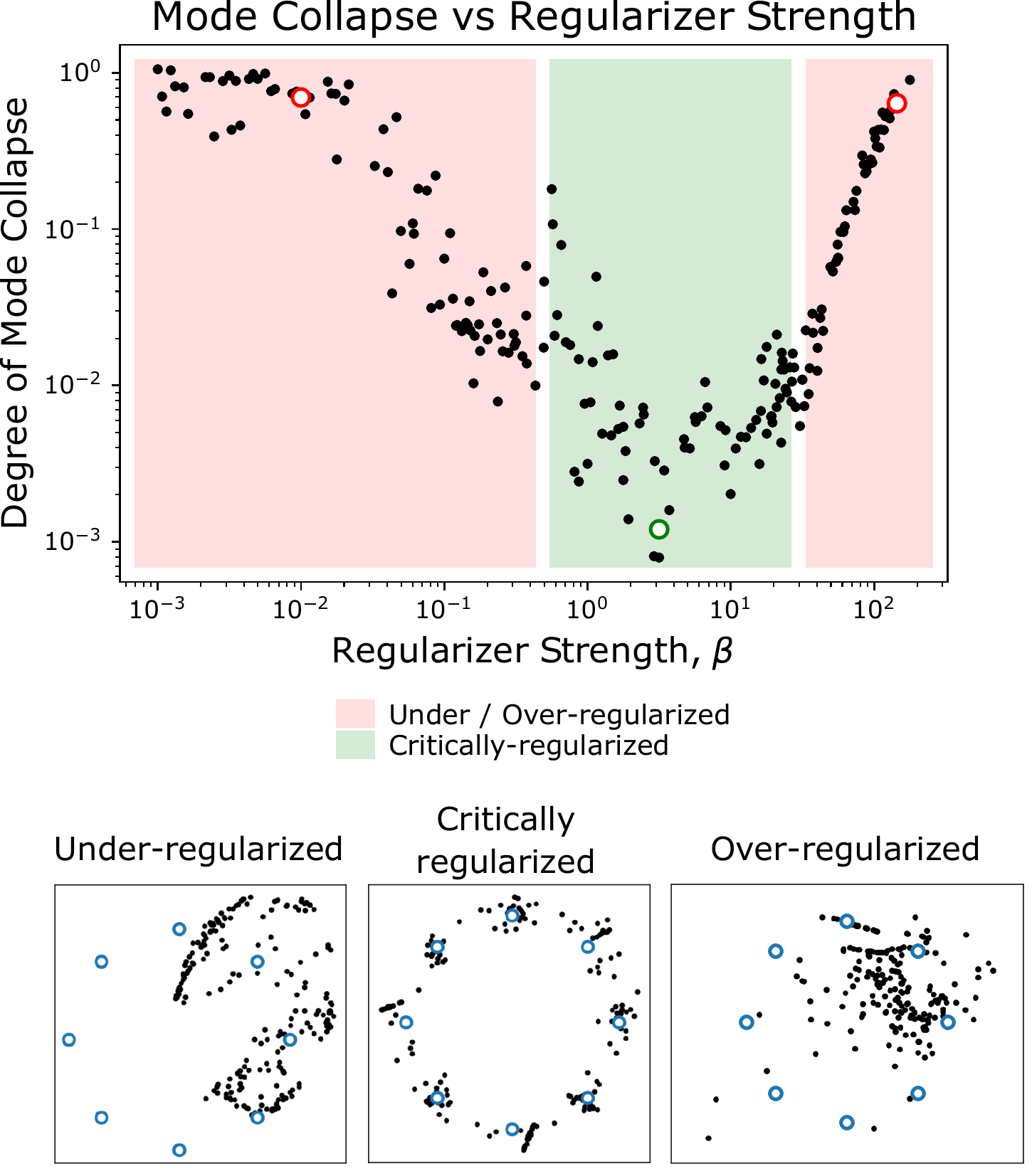}
    \caption{
    \textbf{Critical regularization mitigates mode collapse within a model GAN.}
    Upper: each point represents an experiment using a single-hidden layer ReLU discriminator, halted after 1600 iterations (similar to the experiments of Section~\ref{sec:mode_experiments}). 200 generator points are used, with $g_2 / g_1 = 0.275$ and $n_{disc.} = 5$, placing the system within the regime of mode collapse.
    The regularizer, which plays a role similar to that of a damping term, results in a reduction of mode collapse. We observe three regimes analogous to under, critically, and over-damped dynamics.
    Lower: snapshots of sampled generator particle configurations for the three circled points in the upper diagram. Note that only the critically regularized example has converged. 
    }
    \label{fig:reg_dyn}
\end{figure}

Within our setup, we can experiment with the regularization parameter $\beta$. Running model GAN experiments using different $\beta$ values, we note regions corresponding to under and over-regularization, and an intermediate regime of critical regularization. In this regime, convergence is most efficiently achieved (Fig. \ref{fig:reg_dyn}).

\section{Discussion}

In this paper, we consider a model GAN system constructed by incorporating limiting features present within real GANs. 
The generator inputs are taken to be sampled uniformly from a sphere of high dimension.
Additionally, the generator is assumed to be of infinite width and to have a static NTK such that given inputs of fixed magnitude, the NTK is a function only of their dot product ($\Gamma(z, z') = \Gamma(z \cdot z')$, as is the case in infinite-width ReLU networks).
We also modify the training procedure, using a single fixed mini-batch of generator seeds throughout training.
Under these approximations, the outputs of an infinite-width generator may be represented as a cloud of particles, whose velocities are coupled through the generator network's NTK. Further, due to the nature of the assumed NTK and the high dimensional inputs, we argue that this coarse-grained NTK may be characterized using just two values.

Despite the simplicity of our model GAN system, we observe that it is able to exhibit the defining symptoms of mode collapse – generator outputs fail to become diverse. Indeed, the simplified particle-based setting allows for physical interpretation of the phenomenon through competition between the local gradient experienced by each particle, and the average discriminator gradient experienced by the cloud as a whole. When the latter dominates over the former, the particle cloud fails to split and hence cannot cover a target diversity of modes.  
From this physically motivated effective description, we are able to connect the ratio of the two effective NTK values to the occurrence or avoidance of mode collapse. 

Because the generator NTK values within the model GAN setup can be easily modified, our framework makes it possible to study learning dynamics over a broad range of generators.
Using simple discriminators with a single hidden layer, we investigated the onset of mode collapse as a function of the NTK parameters and the relative training rates of the discriminator and generator. 
We were able to identify power-law and exponential relationships, and explain their presence by drawing a connection to the frequency principle; frequency-dependent learning rates alone suffice to explain the shape of the transition boundary. To our knowledge, this is the first time that the principle has been observed in the context of GANs. 
As a consequence, for a given NTK matrix, mode collapse can be avoided by allowing sufficient time for the discriminator to learn finer features characterizing a multi-modal target distribution. 

Would it be possible to reduce the training time while avoiding mode collapse?
We have experimented with a regularizer designed to reduce mode collapse in real GANs, the effect of which is to dampen the velocity of the generator parameters during training. 
Despite the fact that our model contains no generator parameters, and focuses instead on the dynamics of the outputs, we show how it is possible to adapt such a regularizer to our particle-based setting.
We demonstrate that, in our effective model too, such a regularizer can encourage smooth paths to convergence.
Importantly, an analogy to a damped oscillator, made clear through examples, enables us to identify regimes analogous to over-damping, under-damping, and critical damping. Intermediate regularization strengths would allow most efficient convergence, suggesting 
``critical regularization" as a potential means to cure mode collapse and shorten training time.

The problem of understanding GAN convergence is complex.
By essentializing key features of real GANs, we have probed GAN failure in a more physically interpretable setting, which allows for extensive experimentation.
However, the model's assumptions also suggest directions of future work in studying deviations from these limiting approximations, and in mapping the lessons learned to more realistic GAN settings. 

We have, for instance, assumed a time-independent NTK with uniform values throughout data-space. In reality, however, for networks of finite width, the NTK evolves throughout training. Indeed, such dynamic corrections may be studied order by order (in $1 / \text{network width}$) \cite{deep_learning_theory_book} and incorporated into a more complete analysis. 
An NTK function which develops spatial features during training 
(that is, an NTK defined in data-space, $\Gamma(X, X')$)
 might yield dynamics showing closer parallels to flocking, in which individual birds look at spatial neighbors to update their velocities \cite{flock, toner_long-range_1995}.
 
We have also replaced the distribution of NTK values (comprising the evaluations of $\Gamma(z, z)$ and $\Gamma(z \neq z')$ within a mini-batch) with just two numbers: $g_1$ and $g_2$. 
 By including some variance in the diagonal and off-diagonal NTK values, as is present in realistic settings of finite latent-space dimension (see, for example, Fig.~\ref{fig:NTK_distribution}), future work might broaden the scope of the noted universality.

 Finally, we have obtained our results using a modified training algorithm in which only one mini-batch of seeds is used throughout training. In order to extrapolate the lessons learned to a more realistic setting, we would like to better understand the implications of our results for contexts in which mini-batches of seeds $\{ z\}$ are continually resampled.

\section{Acknowledgments}
During the preparation of this work, Steven Durr was supported by funds from the Bhaumik Institute for Theoretical Physics at UCLA. This work used computational and storage services associated with the Hoffman2 Cluster hosted by the UCLA Institute for Digital Research and Education. Shenshen Wang is grateful for support from an NSF CAREER Award (Grant No. PHY-2146581).

\newpage
\appendix

\section{\label{app:exp_details} NTK Sweep Experimental Details}

The experiments of Section~\ref{sec:mode_experiments} are performed using the following protocol:

\begin{itemize}
    \item Generator: A collection of 200 two-dimensional points initialized as a gaussian centered at $(\sqrt{2}, \sqrt{2})$, with standard deviation $.1$.

    \item Discriminator: A single hidden layer neural network of width 2048,
    \begin{equation}
        D(x) = \sqrt{\frac{2}{\text{width}}}a_i \sigma(w^j_i x_j + b_i).
    \end{equation}
    Experiments were run using both ReLU and Tanh activation functions.
    At initialization, we take
    $ a_i, \ w^j_i \sim \mathcal{N}(0, \sigma^2 = 1)$ and $b_i \sim \mathcal{N}(0, \sigma^2 = 9)$.
\end{itemize}

To understand the relationship between the generator NTK and training rate, we vary both the discriminator learning rate, $\alpha_D$, and the discriminator updates per iteration, $n_{disc.}$ (Algorithm \ref{alg:gan_training_2}). We then examine the degree of mode collapse after some fixed number of training iterations.

The results of experiments which sweep over $n_{disc.}$ are described in the main text (Section~\ref{sec:mode_experiments}), and the results of those varying $\alpha_D$ are given in Section~\ref{sec:alpha_experiments}. Both reflect the same pattern: generally, for larger $g_2 / g_1$, the discriminator requires more `time' (larger $\alpha_D$ or greater $n_{disc.}$) in order for the adversarial dynamic to overcome mode collapse. Additionally, power-law and exponential mode collapse boundaries are for the ReLU and Tanh discriminators, respectively. 

In our experiments, as we vary $g_2/g_1$, we take care to control for the overall effect that the NTK has on total velocity.
For example, if all points were initialized at the same location, $X$, then their velocities would obey
\begin{align*}
    \frac{d X_a}{dt} &= \alpha_G \frac{1}{N}\sum_j \Gamma_{a, b} \nabla D(X_b) \\
    &= \alpha_G \frac{\nabla D(X)}{N} (g_1 + g_2 (N-1))
\end{align*}
To control for this effect, as we vary $g_2/g_1$ we maintain $g_1 + g_2 (N-1) = \text{const.}$ We take $g_1 + g_2 (N-1)=N$, so that $g_2 / g_1 = 0 \implies g_2=0, \ g_1 = N$, and $g_2 / g_1 = 1 \implies g_2= g_1 = 1$. Since in our experiments, points are initialized in a tight distribution (with $\sigma=.1$), we believe this allows us to meaningfully compare the effect of different generator NTKs.

The NTK values written in Eq.~(\ref{eq:data_space_dynamics_2}) can be thought of as elements of an NTK Gram Matrix \cite{freeze_chaos_ntk}. Our choice of ($g_1$, $g_2$) normalization is then equivalent to fixing the eigenvalue of the constant mode to $\lambda_{const} = N$ for all $\phi$. All other eigenvalues then equal
$$
\frac{N}{N-1}\left(\frac{N}{(N-1)\phi + 1} -1 \right).
$$

From this perspective, as $\phi$ grows, the constant mode of the NTK Gram matrix dominates, a fact which has previously been associated with the presence of mode collapse \cite{freeze_chaos_ntk}.

\subsection{Discriminator Learning Rate Experiments}
\label{sec:alpha_experiments}
Rather than varying $n_{disc.}$, we run experiments which vary the training rate of the discriminator learning rate, $\alpha_D$. Apart from this difference, the experiments performed are identical to those of Section~\ref{sec:mode_experiments}.

\begin{figure}[ht]
    \centering
    \includegraphics[width=.5\textwidth]{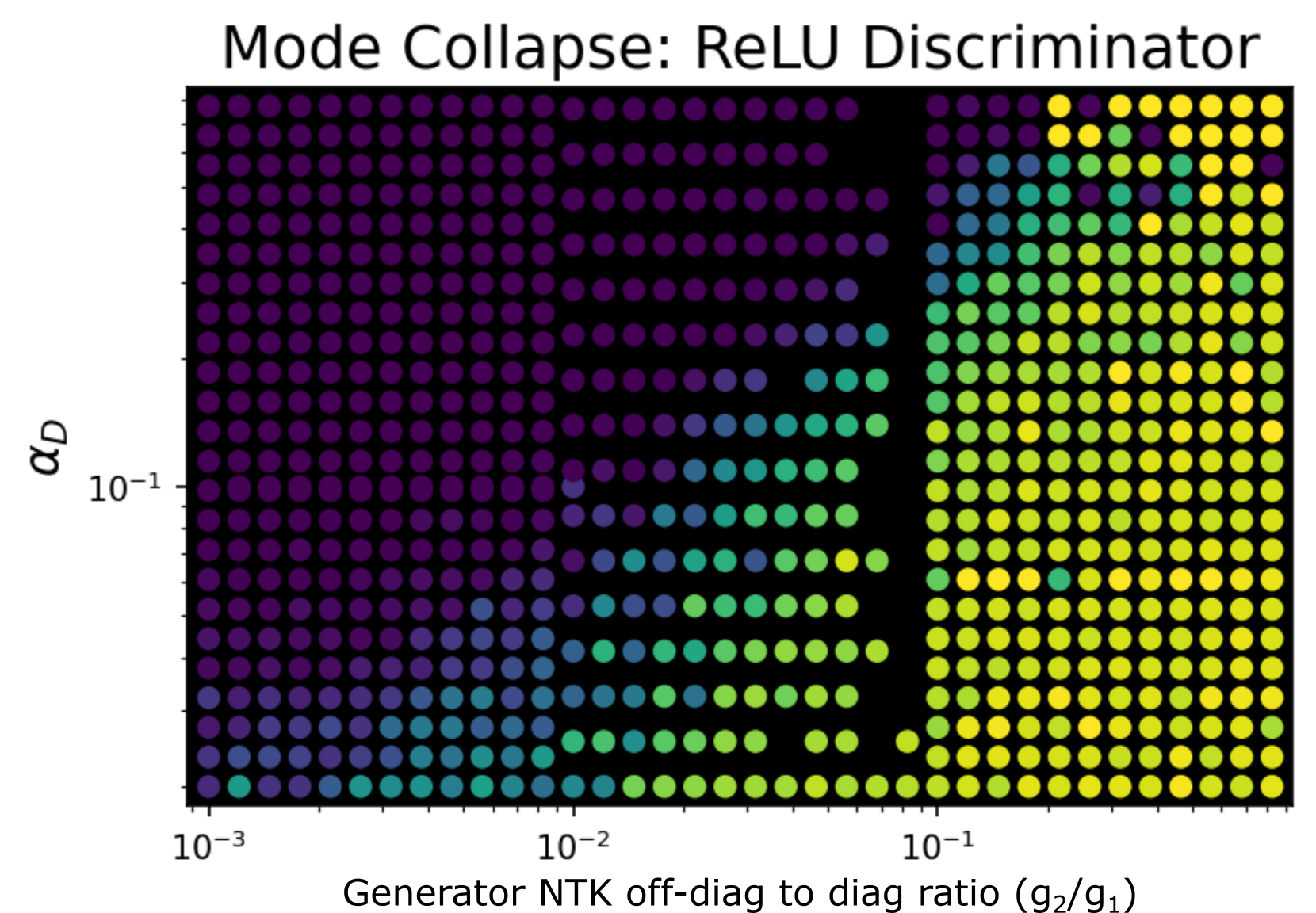}
    \caption{
    \textbf{A power-law mode collapse threshold is found for ReLU discriminators.}
    Mode collapse is depicted as a function of $\alpha_D$ and the ratio $g_2 / g_1$ using a ReLU discriminator. Mode collapse data is averaged over the iterations, $4000 \pm n \times 20$, with $n=0, 1, 2, 3$. On the log-log plot a linear threshold is observed.
    }
    \label{fig:dstep_relu}
\end{figure}

\begin{figure}[htb]
    \centering
    \includegraphics[width=.5\textwidth]{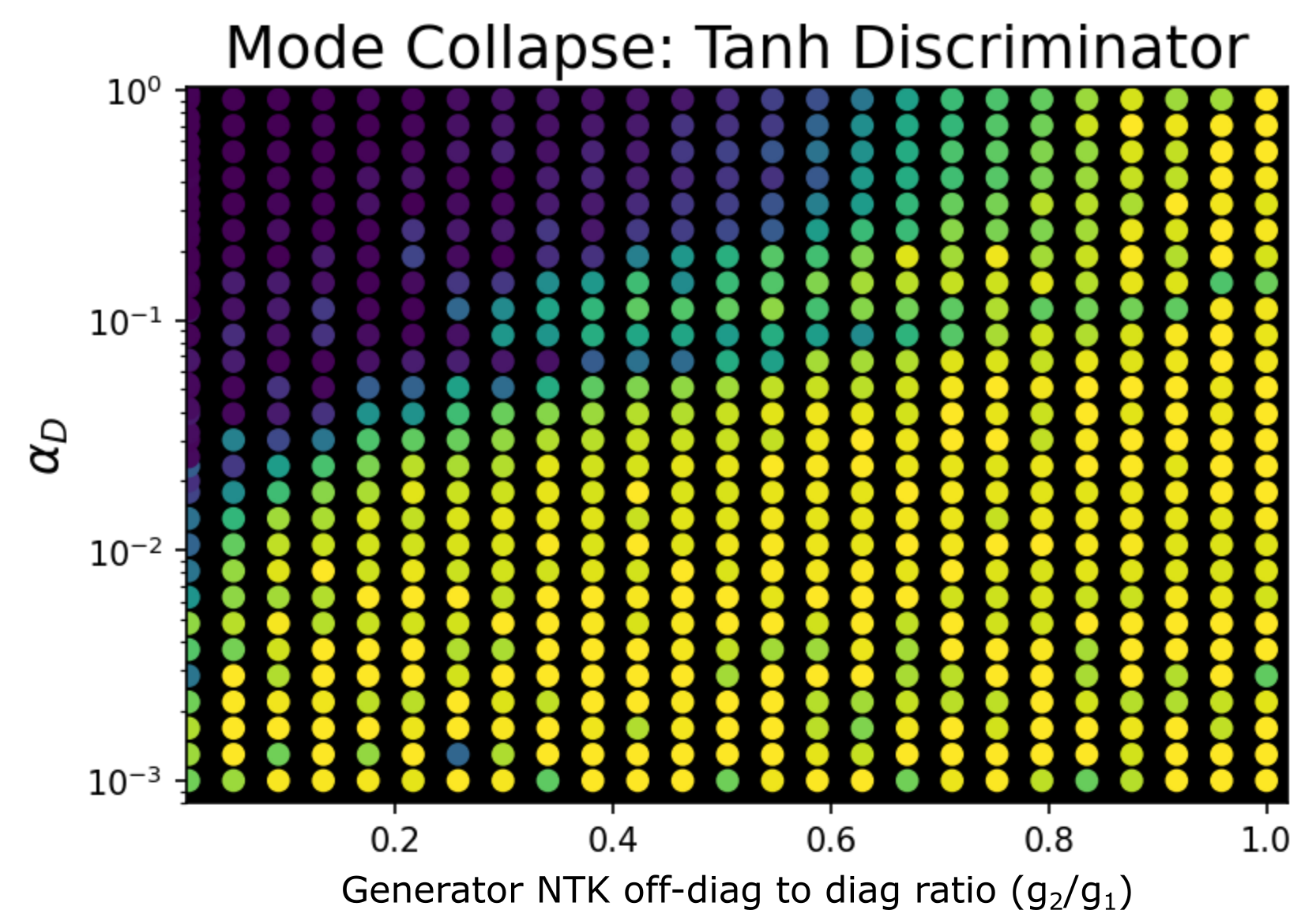}
    \caption{
     \textbf{An exponential mode collapse threshold is found when using Tanh discriminators.}
     Mode collapse as shown as a function of discriminator learning rate, $\alpha_D$, and $g_2 / g_1$. As in Fig.~\ref{fig:dstep_relu}, mode collapse is averaged about iteration 4000. On a log-linear plot, a broadly linear threshold is observed, indicating an exponential transition.
    }
    \label{fig:dstep_tanh}
\end{figure}

Using a ReLU discriminator, a roughly power-law transition boundary is found (shown in Fig.~\ref{fig:dstep_relu}). A Tanh discriminator, shows an exponential boundary (shown in Fig.~\ref{fig:dstep_tanh}). As in the case of sweeps over $n_{disc}$, this matches the expected frequency learning rate, $\gamma(k)$, of the respective networks \cite{frequency_vs_training, convergence_vs_freq, linear_f_principle}.


\section{Precision Scaling Arguments}
\label{section:precision_scaling_argument}

\begin{figure}[ht]
    \centering
    \includegraphics[width=.5\textwidth]{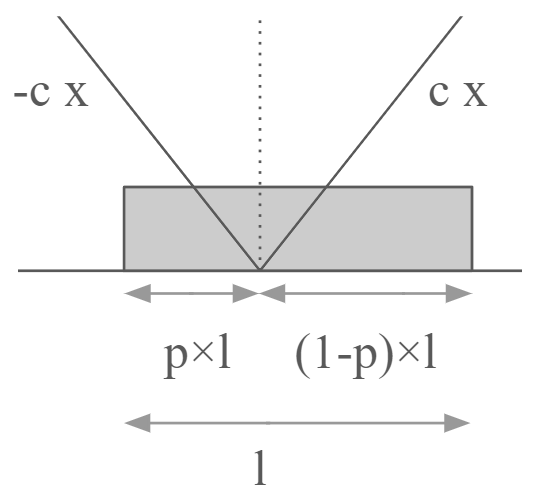}
    \caption{
    \textbf{A schematic depicts a hypothetical discriminator and generator distribution.}
    The generator's uniform distribution of length $l$, depicted shifted a distance $p l/2$, $p\in[0,1]$, to the right of the origin. The discriminator of the form $c|x|$ is superimposed.
    }
    \label{fig:mech}
\end{figure}

The correspondence between ReLU networks and power-law boundaries (Fig.~\ref{fig:relu_data}), and Tanh networks and exponential boundaries (Fig.~\ref{fig:tanh_data}), can be interpreted by considering the spatial precision required for a discriminator to split apart a collection of particles uniformly distributed within a one-dimension region. In Fig.~\ref{fig:mech}, we depict a hypothetical discriminator (defined by $D(x) = c |x|$), and a distribution of generator points uniformly distributed within $[-p l, (1-p)l]$ (depicted as a shaded region). The placement of discriminator's minimum with respect to the center of the generator distribution is determined by $p$, with $p=1/2$ placing its minimum directly at the center, and $p=0$ completely shifting the distribution to the right side of $D(x)$.

Under this setup, the velocities of the points to the left and right of the minimum of $D(x)$ will be
$$
v_L = c \cdot \left( -\frac{(g_1 - g_2)}{N} + g_2 (1-2p)\right),
$$
and
$$
v_R = c \cdot \left( \frac{(g_1 - g_2)}{N} + g_2 (1-2p)\right).
$$

These velocities satisfy
$$
v_L<0 \implies p > \frac{1}{2} - \frac{g_1 - g_2}{2 g_2 N},
$$
$$
v_R>0 \implies p < \frac{1}{2} + \frac{g_1 - g_2}{2 g_2 N}.
$$

In order to `split' the points, and ensure that the particles on each side have opposing velocities, we require
$$
\left(\frac{1}{2} -\frac{g_1 }{2 N g_2} + \frac{1}{2N}\right) < p < \left(\frac{1}{2} +\frac{g_1 }{2 N g_2} - \frac{1}{2N}\right).
$$

This range of $p$ indicates that for the discriminator to be able to split apart the distribution, we require the discriminator's minimum to be near the center of the generator distribution, with a \textit{spatial precision} of order
$$
l \frac{g_1-g_2}{N g_2}.
$$
Here if we take $g_1/g_2$ to be large, then the relevant frequency corresponding to this spacial precision is roughly
\begin{equation}\label{eq:freq}
    k \sim \frac{N g_2}{l g_1}.
\end{equation}

\section{Supporting the F-Principle Mechanism}
\label{app:fprince_support}

\begin{figure}[h]
    \includegraphics[width=.5\textwidth]{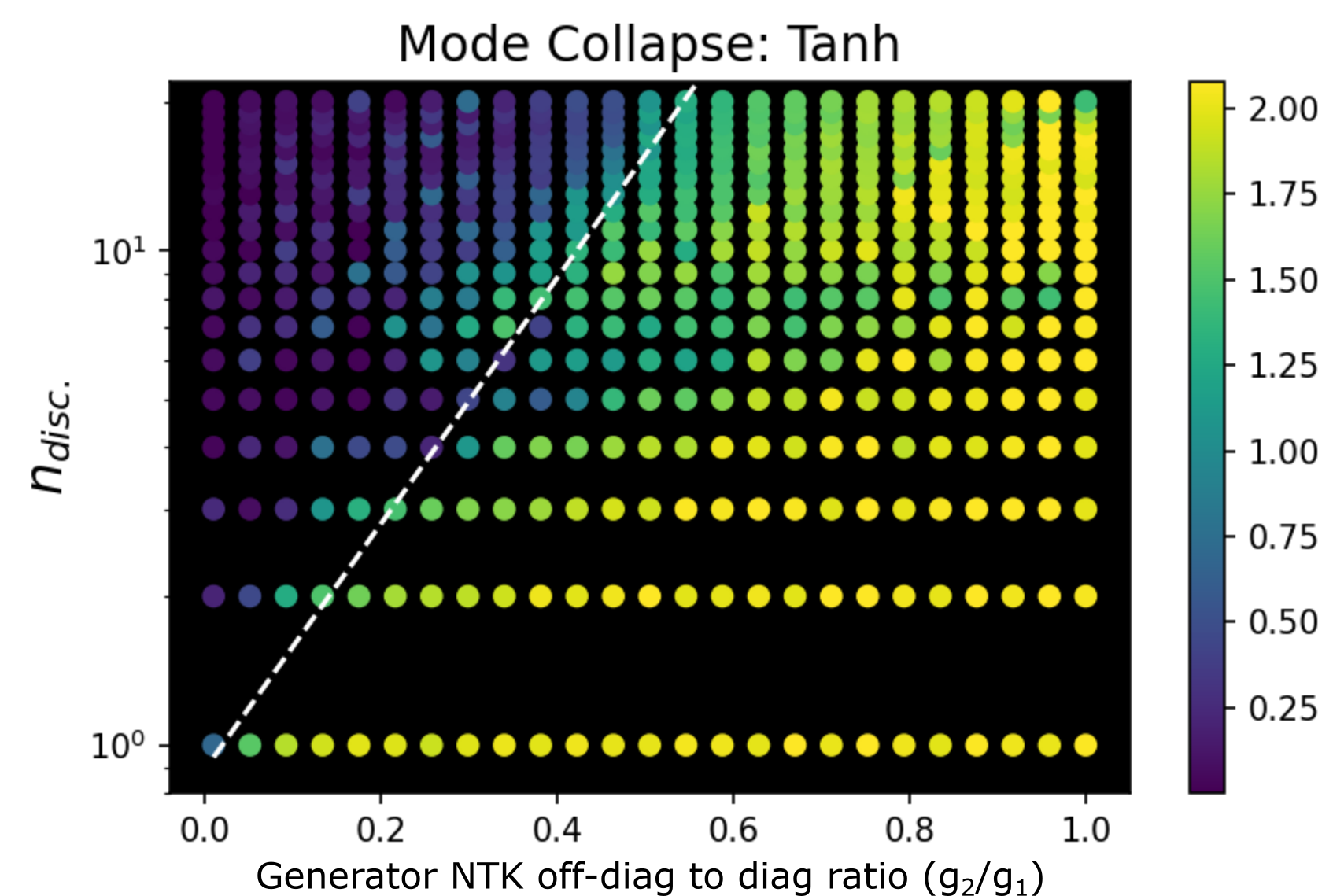}
    
    \caption{
    \textbf{Phase diagram using a Tanh discriminator.}
    Results are depicted as in Fig.~\ref{fig:relu_data}, although experimental results are taken after 2500 training iterations.
    A dashed line is fit to the transition, highlighting a roughly exponential phase boundary, and differing from the power-law boundary observed in Fig.~\ref{fig:relu_data}.
    }
    \label{fig:tanh_data}
    \end{figure}

Analogous to the mode collapse experiments using a ReLU discriminator (Section~\ref{sec:mode_experiments}), experiments were also performed employing Tanh discriminators (Fig.~\ref{fig:tanh_data}). Here, a roughly exponential phase boundary was found, appearing to match the predicted exponential frequency learning rate, $\gamma(k)$ \cite{linear_f_principle}. 

Our physically motivated mechanism for the transition (described in Section~\ref{interpretation}) makes use of the so-called frequency principle within neural networks to explain the shape of the phase boundary. 
In light of the stark contrast between the shapes of the ReLU and Tanh boundaries (shown in Figures \ref{fig:relu_data}, \ref{fig:tanh_data}), which match the differences in their respective frequency learning rates \cite{linear_f_principle}, this connection appears very plausible. 

We would like, however, to ensure that such a frequency relationship is \textit{sufficient} on its own to create such power-law and exponential phase boundaries, since it is conceivable that some other property of the networks is responsible.

We note that our discriminators are very wide networks with a single hidden layer. In this large-width limit, it is expected to be approximately linear in parameters during training \cite{wide_networks_are_linear}. Additionally, the networks in question are known to obey a given frequency principle. We therefore define a new discriminator which has these precise properties alone, and rerun the same experiment to observe the resulting phase boundary. If the same power-law and exponential phase boundaries are found, we can be much more confident in this connection.

We define,
\begin{align}
\label{eq:fourier_disc}
    D(x) &= \sum_k D_k(x), \\
    D_k(x) &= w^{(1)}_k \sin(k \cdot x) + w^{(2)}_k \cos(k \cdot x),
\end{align}
where $k = (k_1, k_2)$ and $k_i$ are taken from 25 values of equal logarithmic spacing from [.01, 20], as well as the negatives of these values. $w^{(i)}_k$ are the weights of the model.

During training, we follow the routine of Algorithm \ref{alg:fourier_alg}, a modification of Algorithm \ref{alg:gan_training_2}, in which each $w^{(i)}_k$ is updated with a rate proportional to the value of a function, $\gamma(k)$. We then plug in power-law and exponential $\gamma(k)$ functions by hand, and run the same experiments performed in Section~\ref{sec:mode_experiments}. The power-law and exponential $\gamma(k)$ functions are defined below \footnote{These functions were obtained by experimenting with the $\gamma(k)$ functions corresponding to real neural networks, and finding approximate matches for our sum-of-Fourier-mode discriminators.}:
\begin{align}
    \gamma_{\text{pow.}}(k) &= \min(10^3, |k|^{-3}), \\
    \gamma_{\text{exp.}}(k) &= 668.8 \cdot \exp(-2.05 \cdot |k|).
\end{align}

\begin{algorithm}
\caption{The model-GAN training algorithm, with a Fourier-Discriminator and a frequency-dependent learning rate.}\label{alg:fourier_alg}
\begin{algorithmic}
\For{iteration number}
    \For{$n_{disc.}$}
        \State $\bullet$ Sample $N$ data-points, $\{ x_i \}$, from the 8-Gaussian distribution.
        
        \State$\bullet$  Compute 
        $$
        \mathcal{L}^{(N)} = \frac{1}{N} \sum_{a=1}^N D(X_a) - \frac{1}{N} \sum_{i=1}^N D(x_i) + \frac{\lambda}{2}\sum_k \left( (w^{(1)}_k)^2 + (w^{(2)}_k)^2\right)
        $$
        \State and update discriminator parameters by ascending its stochastic gradient 
        $$
        w^{(i)}_k \gets w^{(i)}_k + \alpha_D \ \gamma(k) \ \nabla_{w^{(i)}_k} \mathcal{L}_D^{(N)}
        $$
    \EndFor
    \State $\bullet$ update $X_a$ according to Eq.~(\ref{eq:data_space_dynamics_1})
    $$
    X_a \gets X_a + \alpha_G \ \frac{1}{N} \sum^N_b \Gamma_{a, b} \nabla_x D(X_b)
    $$
\EndFor
\end{algorithmic}
\end{algorithm}

Our new routine essentializes the properties of the wide ReLU and Tanh discriminators by being linear in the parameters and explicitly learning frequency $k$ features with a rate $\gamma(k)$. 

\begin{figure}[htb]
    \centering
    \includegraphics[width=.5\textwidth]{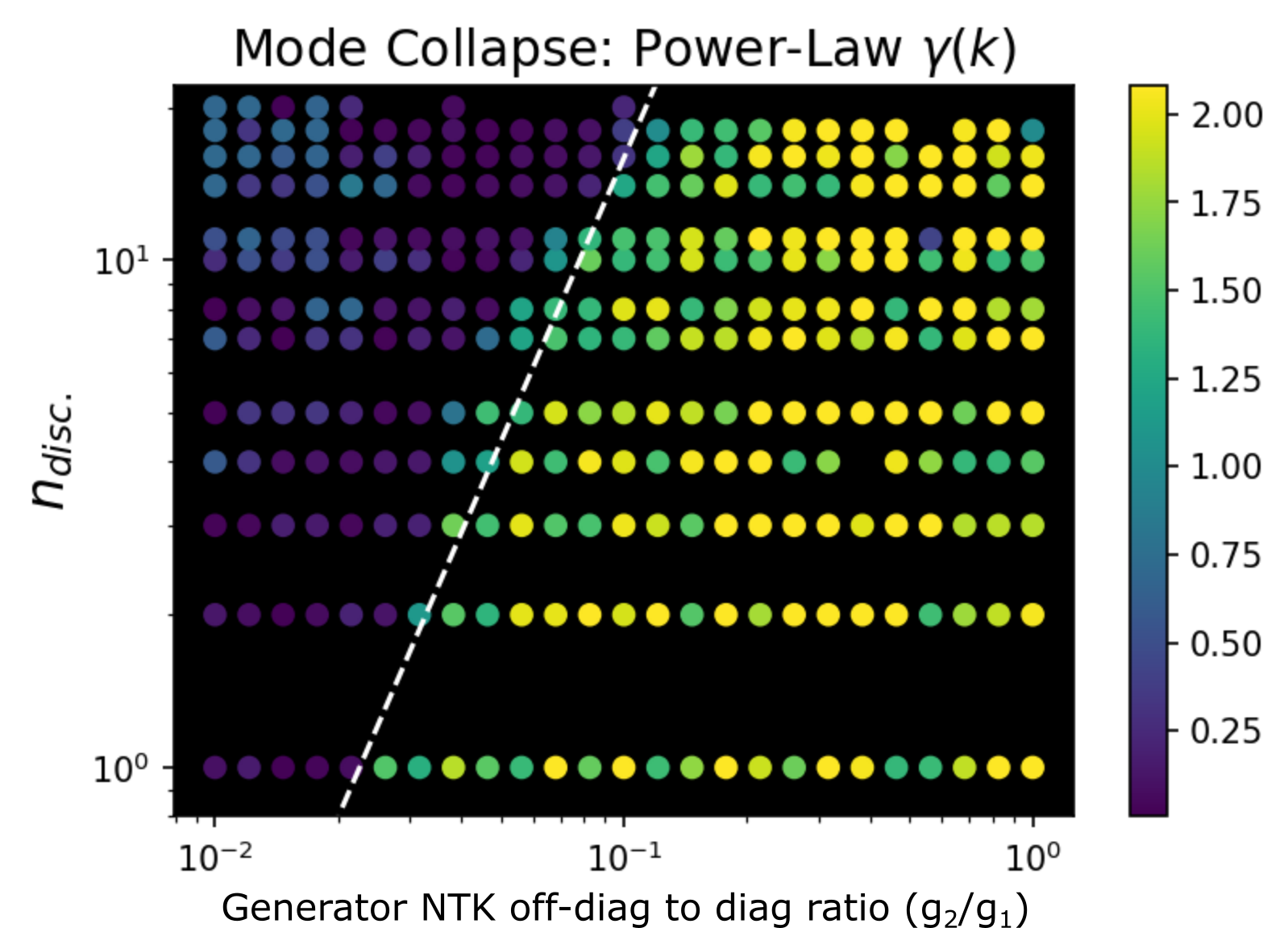}
    \caption{
    \textbf{Fourier discriminators (Eq.~\ref{eq:fourier_disc}) with power-law $\gamma(k)$ have a power-law mode collapse transition.}
    A scatter plot depicts the transition for a power-law $\gamma(k)$ after 3000 steps. Brighter points indicate mode collapse, and darker points indicate convergence. An extremely clear power-law boundary is found here, with a slope of $\approx 4.90$.
    }
    \label{fig:power_fourier}
\end{figure}

\begin{figure}[htb]
    \centering
    \includegraphics[width=.5\textwidth]{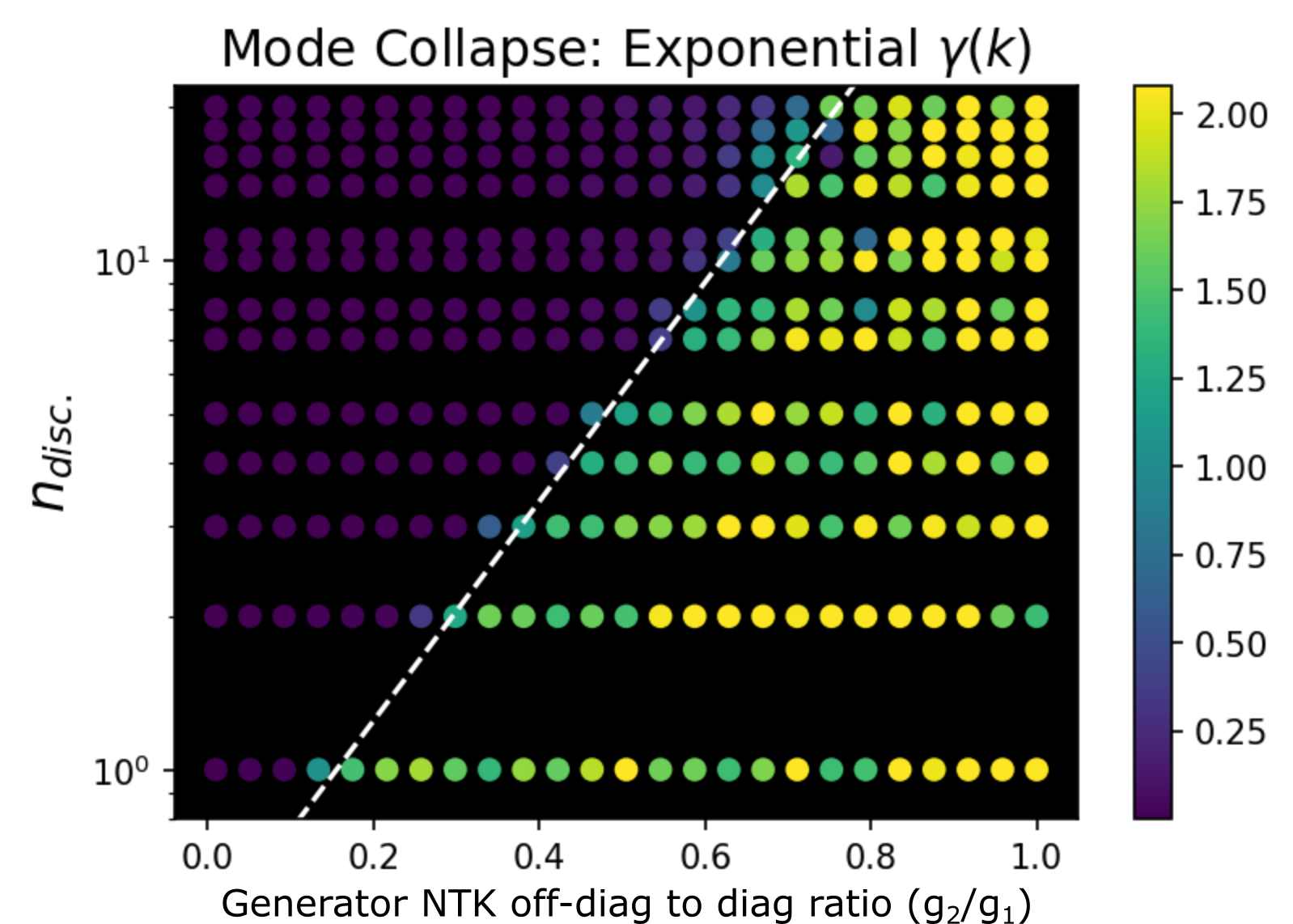}
    \caption{
    \textbf{Fourier discriminators (Eq.~\ref{eq:fourier_disc}) with exponential $\gamma(k)$ show an exponential mode collapse transition.}
    A scatter plot shows the transition for an exponential $\gamma(k)$ after 2000 steps. Brighter points indicate mode collapse, and darker points indicate convergence. Again a clear exponential boundary is found here, having a slope of $\approx 1.86$.
    }
    \label{fig:exp_fourier}
\end{figure}

The results of Figures \ref{fig:power_fourier} and \ref{fig:exp_fourier} show a very clear phase boundary having precisely the power-law and exponential behavior, respectively. This indicates that a frequency dependant learning rate is sufficient to produce the type of phase boundary we previously observed, and lends credence to the connection drawn between the onset of mode collapse, and the frequency principle of the discriminator network.

We do, however, emphasize that our simple explanation of the threshold shape (Section~\ref{interpretation}) is likely incomplete. In particular, the assumption of $g_1 / g_2 \gg 1$ breaks down within Fig.~\ref{fig:exp_fourier}, and yet the depicted transition remains essentially linear (exponential) even to $g_2 / g_1 \approx .7$.
Rather, our description outlines a plausible causal connection, from which a more general explanation might be obtained.

\section{Generator Distributions across the Transition}
\label{app:sample_dists}

To visualize the behavior of the generator points through the transition, here we plot the generator distributions for different $g_2 / g_1$ values given a fixed $n_{disc.}$. This uses a ReLU discriminator, and the outputs of the experiment performed in Section~\ref{sec:mode_experiments}. 

Taking $n_{disc.}=6$, the transition here occurrs roughly at $g_2 / g_1 = 0.06$ (see the transition depicted in Fig.~\ref{fig:dstep_relu}). We therefore show plots from below and above this value of $g_2 / g_1$. 

\begin{figure}[ht] 
\includegraphics[width=.5\textwidth]{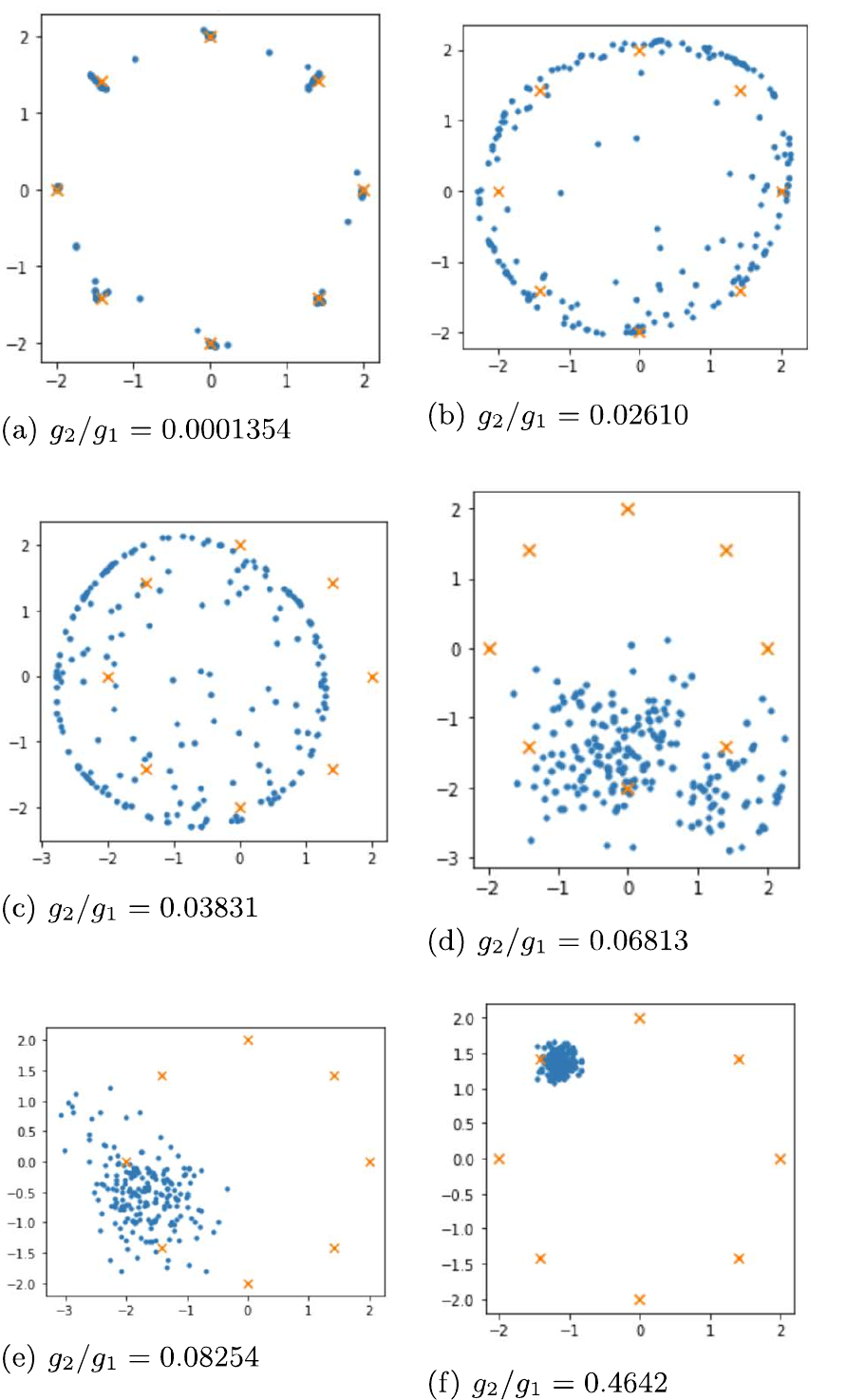}

\caption{
\textbf{Generator particles overcome mode collapse and converge below the $g_2 / g_1$ boundary.}
Generator outputs are shown after 3000 steps. Note the full convergence for small $g_2/g_1$, while for $g_2/g_1>0.06$ the generator fails to converge.} \label{fig:final_gens}
\end{figure}

\begin{figure}[ht]
\includegraphics[width=.5\textwidth]{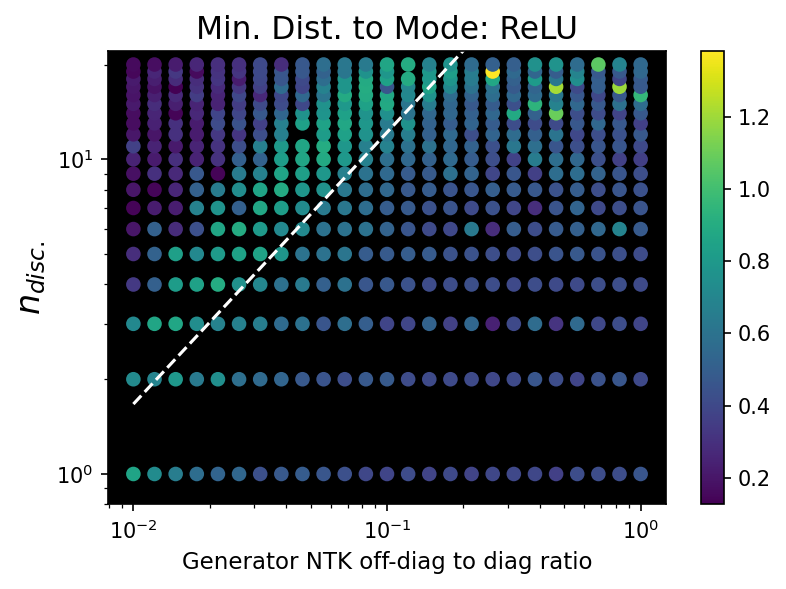}

\caption{
\textbf{Euclidean distance to Nearest Mode drops above (convergence) and below (mode collapse) the mode collapse transition.}
Above the phase boundary, points are close to modes due to convergence. Below the transition, points are close to modes due to mode collapse. We observe an increase near the phase transition, indicating that the initially tight clusters of particles have broken apart, and the process of convergence has begun. This behavior is somewhat apparent in the log-likelihood plotted in Fig.~\ref{fig:relu_loglikelihood_data}.}
\label{fig:relumindist}
\end{figure}

The distribution of generator particles across the mode collapse phase boundary can also be seen through the average (Euclidean) distance to the nearest mode, shown in Fig.~\ref{fig:relumindist}. Below the transition, points are tightly focused, oscillate from mode to mode, and are therefore relatively close to the modes. Far above the transition, the distance to the nearest mode is very small, however this is now due to convergence. Between these two phases, the particles have spread apart. They have begun the process of convergence, and therefore have a larger distance to the nearest mode. 

\section{Regularization and Critical Damping}
\label{app:dirac_gan_reg}

To understand the physical meaning of the regularizer implemented in Section~\ref{sec:critreg}, we can consider its effect on a so-called Dirac-GAN \cite{which_training_methods}. Here, our generator's implicit distribution is simply a Dirac delta focused at $\theta$, with an output given by 
$
G_\theta(z) = \theta
$, a data-distribution focused at $0$, $\delta(x)$, 
and a discriminator defined by 
$
D_\phi(X) = \phi \cdot X.
$
In this system, equilibrium would correspond to the point $\phi=\theta=0$.

The regularizer in this setup then takes the form:
$$
\beta | \nabla_\theta \phi \cdot \theta|^2 /2 = \beta \phi^2 /2.
$$
In a simultaneous descent/ascent setup, we find that
\begin{align}
    \dot{\theta} &= \nabla_\theta D_\phi(\theta) = \phi, \\
    \dot{\phi} &= -\nabla_\phi (D_\phi(\theta) + \beta \phi^2 /2)= -\theta - \beta\phi.
\end{align}
Diagonalizing, we obtain the eigenvalues of the dynamical matrix: $\left( \beta \pm \sqrt{\beta^2 -4} \right)/2$, giving us critical damping at $\beta=2$.

\begin{figure}[!htbp]
    \centering
    \includegraphics[width=.5\textwidth]{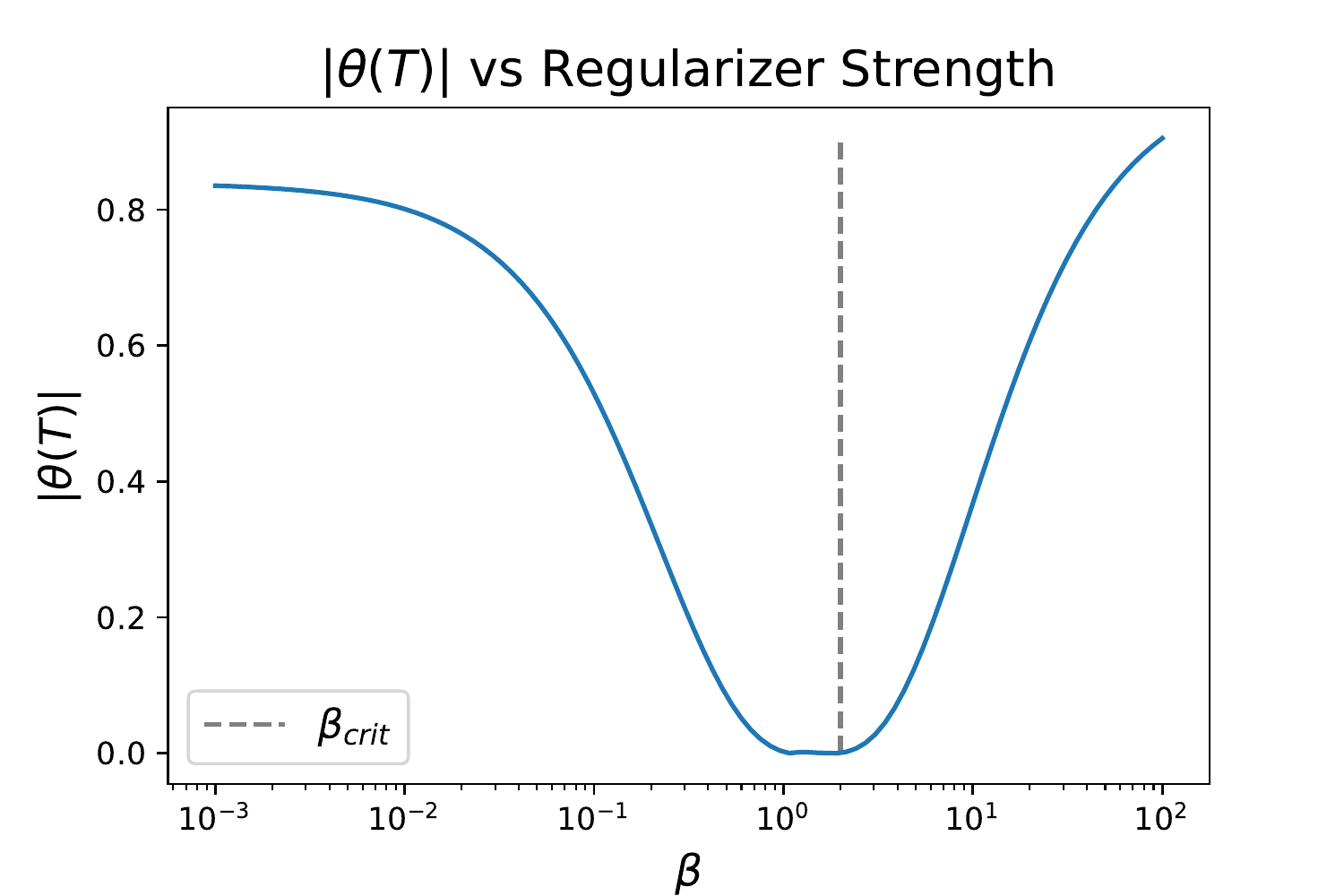}
    \caption{
    \textbf{Critical regularization is demonstrated to result in convergence ($\theta(T)=0$) within a Dirac-GAN.}
    A plot showing the distance after a fixed time, between the Dirac-GAN's generator output, $\theta$, and its equilibrium point, $0$. The system is initialized at $\theta=1$, $\phi=0$, and halted at $T=10$. The critical regularization value, $\beta=2$, is indicated by a vertical dashed line. Note the overdamped, underdamped, and critically damped regions bear a striking resemblance to the three regimes identified in main Fig.~\ref{fig:reg_dyn}.}
    \label{fig:dirac_gan_reg}
\end{figure}

Indeed, if we initialize such a system from $(\theta, \phi) = (1, 0)$ for different $\beta$ values, and observe the value of $|\theta|$ after a set time, $T$ (here we used $T=10$), we obtain Fig.~\ref{fig:dirac_gan_reg}, showing a similar behavior to that found in Fig.~\ref{fig:reg_dyn}. 

\section{Mode Collapse and Regularization for Tanh Discriminators}

\begin{figure}[!htbp]
    \centering
    \includegraphics[width=.5\textwidth]{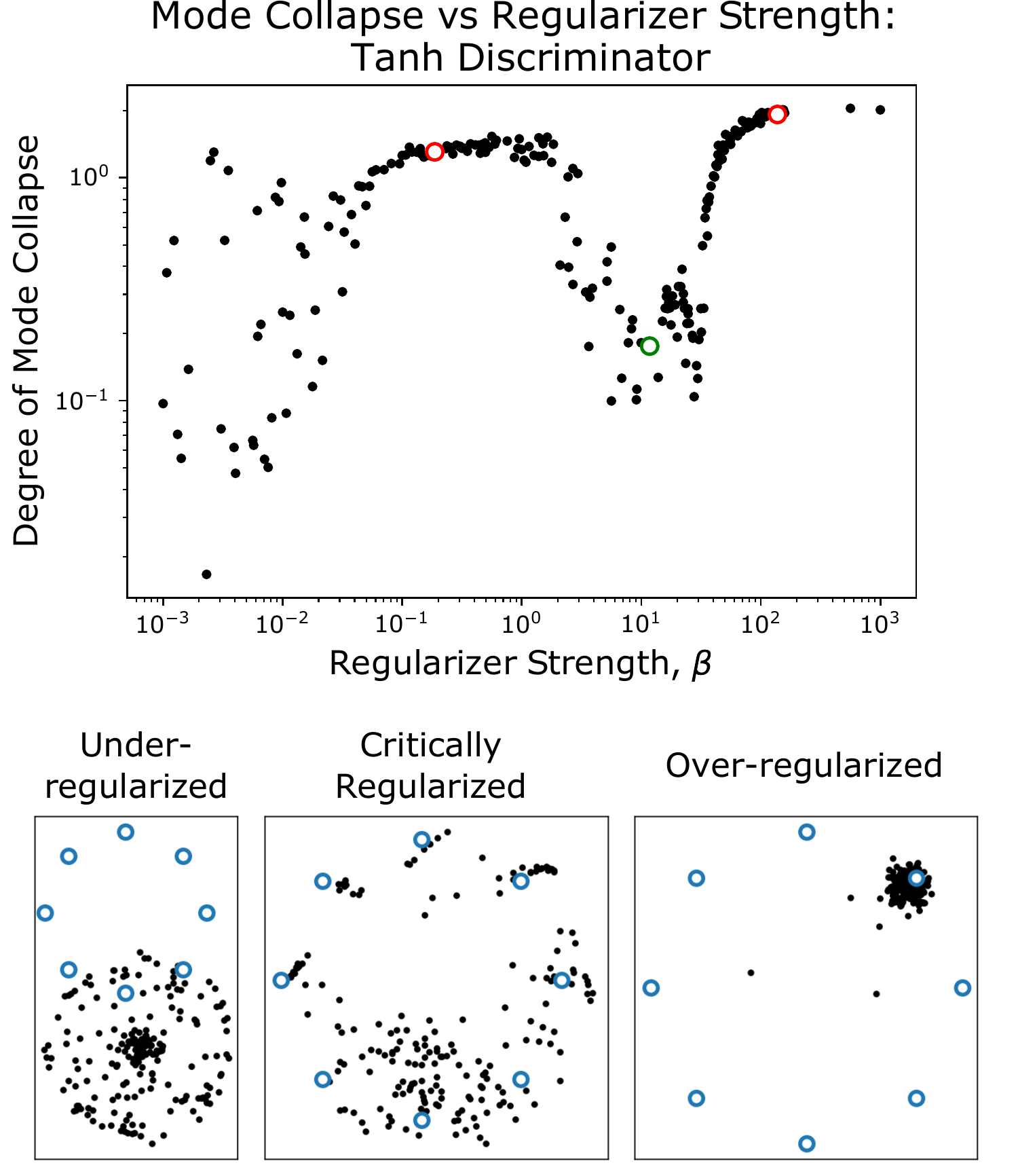}
    \caption{
    \textbf{Regularization regimes are observed using a Tanh Discriminator.}
    A plot of the mode collapse present within a model GAN when trained using a regularizer of strength $\beta$, and a Tanh discriminator. 
    Similar to Fig.~\ref{fig:reg_dyn}, in the upper plot each point represents an experiment halted after 2000 iterations, using a single-hidden layer Tanh discriminator. Here, $g_2/g_1$ is set to $1/5$ and 200 points are used. 
    A clear dip in mode collapse is visible about $\beta \sim 10$. For vanishing $\beta$ values, we observe volatility in mode collapse. Samples from this region are oscillatory, and may oscillate into and out of a more symmetric distribution with respect to the modes. 
    The drop in volatility as $\beta$ is increased reflects the regularizer's influence in encouraging smooth paths to convergence. 
    Sampled generator particle configurations are shown for each of the three circled points, corresponding to under, critically, and over regularized regions.
    }
    \label{fig:reg_dyn_tanh}
\end{figure}

 Using a Tanh discriminator, we compute the degree of mode collapse as regularizer strength $\beta$ is varied. As in Fig.~\ref{fig:reg_dyn}, regimes of over, under, and critical regularization are found (Fig.~\ref{fig:reg_dyn_tanh}).

\section{NTK Evolution During Training}
\label{sec:ntk_evolution}

\begin{figure}[!htbp]
    \centering
    \includegraphics[width=.5\textwidth]{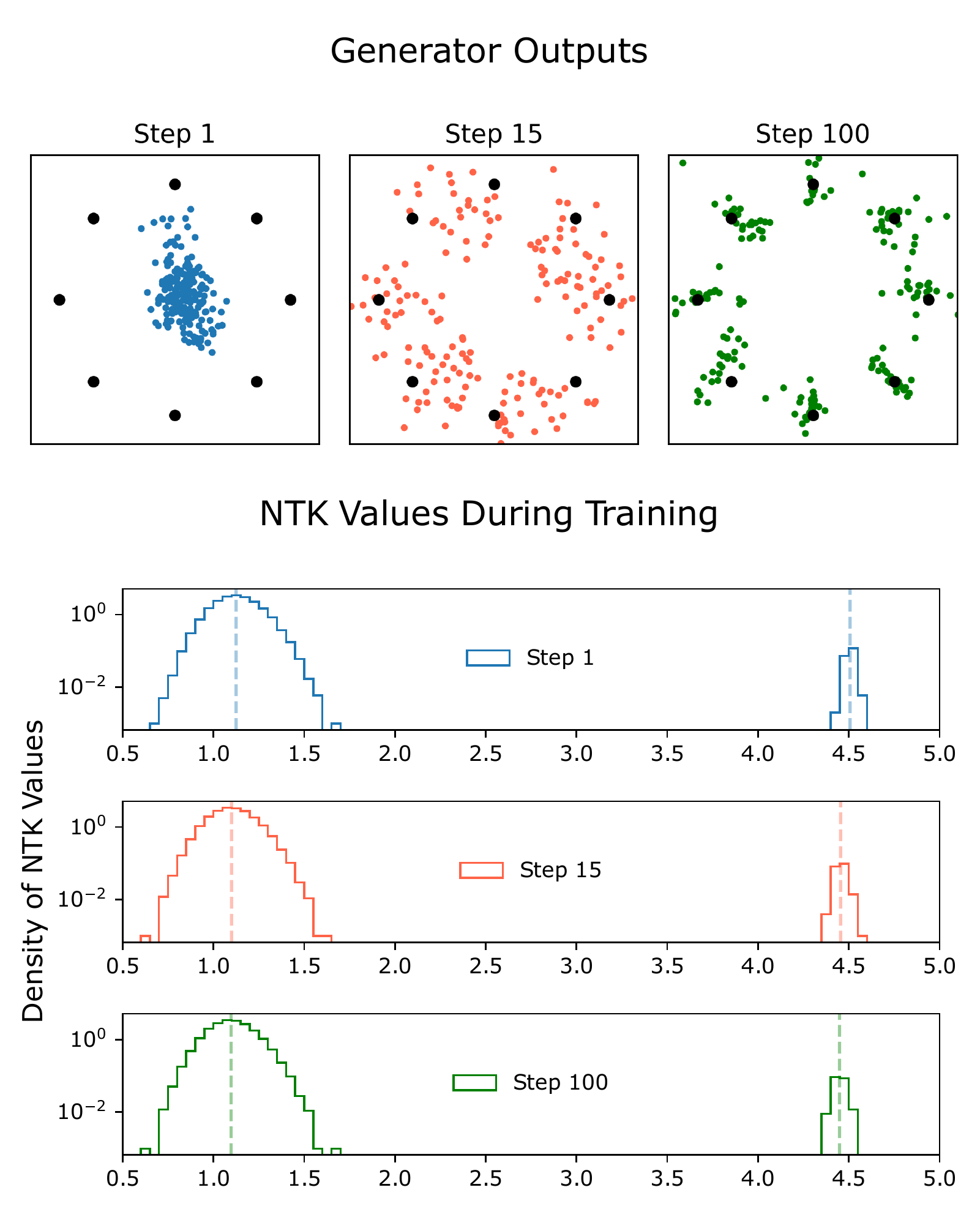}
    \caption{
    \textbf{
    The NTKs of large width generators remain approximately fixed during training, despite the convergence of generator outputs.
    } A single set of seeds is used to train a large width generator (hidden-layer width= $2^{16}$). Generator outputs are shown at three time-slices (above), and the corresponding NTK magnitudes are shown in histograms (below). The medians of the histogram's two peaks are indicated by vertical dashed lines, and roughly correspond to the values of $g_1$ and $g_2$ used in the effective NTK of Eq.~\ref{eq:general_ntk_2}
    }
    \label{fig:gen_ntk_and_outputs}
\end{figure}

In the infinite width limit, the NTK remains fixed during training \cite{jacot_ntk}. An example is shown in Fig.~\ref{fig:gen_ntk_and_outputs}, where despite the convergence of a large-width generator's outputs to the target distribution, its NTK values remain nearly constant. This reflects the assumption we have made in using constant values for $g_1$ and $g_2$ throughout training.

In the upper plot of Fig.~\ref{fig:gen_ntk_and_outputs} is shown the evolution of generator outputs during training. 
At each time-slice, we compute the NTK for each pair of inputs and find that they are very nearly proportional to a $d\times d = 2 \times 2$ identity matrix (reflecting the $\delta_{i, j}$ in Eq.~\ref{eq:general_ntk_2}). 
Below, three histograms show the distributions of NTK magnitudes at each time-slice. The two peaks hardly vary, and correspond to the values of $g_2$ (at $\approx 1.1$) and $g_1$ ($\approx 4.5$) used in Eq.~\ref{eq:general_ntk_2}. 

During training, we use a discriminator and a generator both with a single hidden layer and both using ReLU activations. The generator, expressed,
$$
G^i(x) = \sqrt{\frac{2}{\text{width}}}a_i^j \sigma(w_j^k x_k + b_j),
$$
has a hidden-layer width of $2^{16}$, and parameters initialized according to $w^j_i\sim\mathcal{N}(0, \sigma^2 \approx .046)$, $b_i\sim\mathcal{N}(0, \sigma^2 = 0)$, $a^j_i\sim\mathcal{N}(0, \sigma^2 \approx 2.25)$. Mirroring the training described in the text, throughout training we use only a single set of 200 seeds sampled from a unit sphere in 256 dimensions. The generator is then trained using RMSProp with a learning rate of $10^{-3}$. 

The location of the two peaks within each histogram (which determine $g_1$ and $g_2$) are a function of the network's architecture and the initialization of its parameters. 
For instance, replacing the ReLU activation function, which gives $(g_1, g_2) \approx (4.5, 1.1)$, with an Erf activation yields $(g_1, g_2) \approx (10.6, 5.2)$.  
In general, even when an analytical form of the NTK is available, its value is computed recursively through the layers of the network. Typically no simple closed form is available.

Within certain deep ReLU networks, however, the magnitude of the NTK for orthogonal inputs (corresponding to $g_2$) can be related to the presence of order or chaos within the network \cite{freeze_chaos_ntk}.
Using the notation of \cite{freeze_chaos_ntk, jacot_ntk} \footnote{In other works, the letter $\beta$ is used to scale weights and biases. Here, to avoid conflating this variable with the $\beta$ which scales the gradient regularizer, we instead use $\mu$}, taking inputs from a sphere in $n_0$ dimensions of radius $\sqrt{n_0}$, taking the $l^{th}$ layer to have width $n_l$, and $\sigma(x) = \max(0, x)$, we may write,
\begin{align*}
    \alpha^0(z)&= z\\
    \tilde{\alpha}^{l>0}(z) &\equiv \mu \ b^{(l-1)} + \sqrt{\frac{1-\mu^2}{n_{l-1}}} W^{(l-1)} \alpha^{l-1}(z), \\
    \alpha^{l>0}(z) &\equiv \sigma(\tilde{\alpha}^{l}(z)).
\end{align*}
The output of the neural network function itself is then
$
f_\theta(z) = \tilde{\alpha}^{l}(z),
$
where the parameters, $\theta = \{ (W^{(l)})_i^j, \ b^{(l)}_i \}$, are initialized according to $W^j_i, \ b_i \sim \mathcal{N}(0, 1)$. 

\begin{figure}[hb]
    \centering
    \includegraphics[width=.5\textwidth]{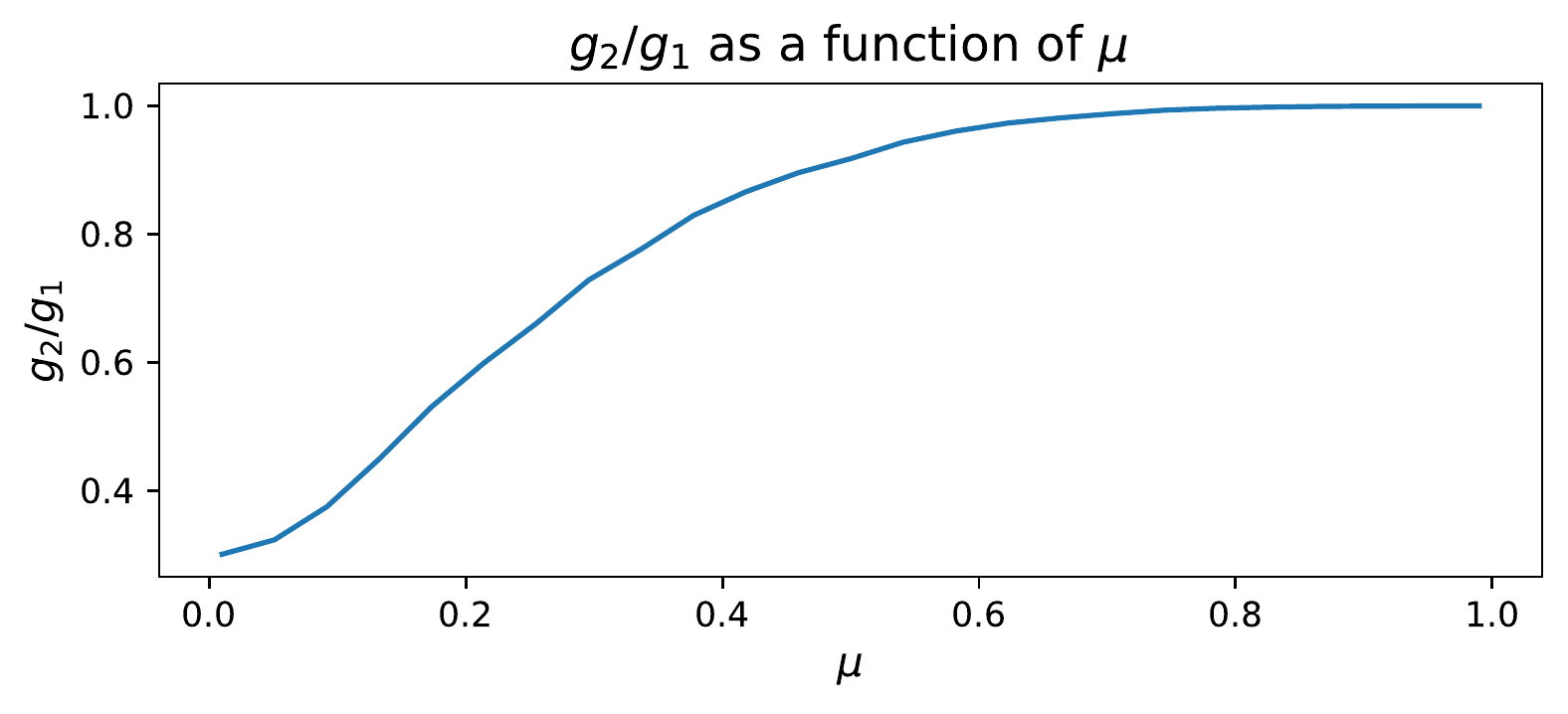}
    \caption{\textbf{Tuning between order and chaos changes the value of $g_2 / g_1$.} 
    More chaotic networks (small $\mu$) have lower values of $g_2 / g_1$ compared to those in the ordered phase ($\mu \approx 1$).}
    \label{fig:g2_g1_ratio}
\end{figure}

Using 6 hidden layers of width $2^{12}$ and $n_0 = 256$, Fig.~\ref{fig:g2_g1_ratio} demonstrates the effect of varying the parameter, $\mu \in [0, 1]$, tuning between networks which are more chaotic and those which are ordered \cite{freeze_chaos_ntk}. 
For values of $\mu$ near 1, the network is expected to be in an ordered phase, and $g_2 / g_1$ approach unity.
Smaller $\mu$ values correspond to networks that are more chaotic, and $g_2 / g_1$ is much lower.

\clearpage

\bibliography{apssamp}

\end{document}